\documentclass[10pt,journal,twocolumn]{IEEEtran}
\newif\ifCLASSOPTIONromanappendices \CLASSOPTIONromanappendicestrue
\usepackage{a0size}
\usepackage{amssymb}
\usepackage{multicol}
\usepackage[english]{babel}
\usepackage{epsfig}
\usepackage{bm}
\usepackage{amsfonts,color,amsthm,amsmath, lscape,graphics}
\usepackage{url}
\usepackage{algorithm}
\usepackage{algpseudocode}

\usepackage[font=footnotesize]{caption}
\usepackage{subcaption}
\usepackage{epstopdf}
\usepackage{algorithm}
\usepackage{algpseudocode}
\usepackage{cite}
\usepackage{relsize}

\DeclareFontFamily{U}{matha}{\hyphenchar\font45}
\DeclareFontShape{U}{matha}{m}{n}{
      <5> <6> <7> <8> <9> <10> gen * matha
      <10.95> matha10 <12> <14.4> <17.28> <20.74> <24.88> matha12
      }{}
\DeclareSymbolFont{matha}{U}{matha}{m}{n}
\DeclareMathSymbol{\odiv}         {2}{matha}{"63}

\usepackage{hyperref}
\usepackage{fix2col}

\renewcommand{\figurename}{Fig.}
\addto\captionsenglish{\renewcommand{\figurename}{Fig.}}

\DeclareMathOperator*{\argmax}{argmax}
\DeclareMathOperator*{\argmin}{argmin}

\newcommand{\bb}{\mathbf{b}}

\newcommand{\bh}{\mathbf{h}}
\newcommand{\bg}{\mathbf{g}}
\newcommand{\bv}{\mathbf{v}}
\newcommand{\bV}{\mathbf{V}}
\newcommand{\bu}{\mathbf{u}}

\newcommand{\bI}{\mathbf{I}}
\newcommand{\bw}{\mathbf{w}}
\newcommand{\bx}{\mathbf{x}}

\newcommand{\by}{\mathbf{y}}
\newcommand{\bz}{\mathbf{z}}
\newcommand{\bA}{\mathbf{A}}

\newcommand{\bW}{\mathbf{W}}

\newcommand{\bn}{\mathbf{n}}

\renewcommand{\frac}{\dfrac}

\newcommand{\Tr}{{\mbox{Tr}}}

\definecolor{myOrange}{rgb}{1,0.5,0}
\definecolor{myGreen}{rgb}{0,0.5,0}
\newcommand{\changeb}[1]{{\color{black}#1}}
\newcommand{\changer}[1]{{\color{black}#1}}
\newcommand{\changeg}[1]{{\color{black}#1}}
\newcommand{\changeW}[1]{{\color{black}#1}}
\newcommand{\changeWW}[1]{{\color{black}#1}}
\newcommand{\changeF}[1]{{\color{black}#1}}
\newcommand{\changeFF}[1]{{\color{black}#1}}
\newcommand{\changelast}[1]{{\color{black}#1}}
\newcommand{\Elemsquare}{\mathbin{\text{$\vcenter{\hbox{\textcircled{\tiny{$2$}}}}$}}}

\begin{document}
\title{Deep Active Learning Approach to Adaptive Beamforming for mmWave Initial Alignment
}

\author{{Foad~Sohrabi},~\IEEEmembership{Member,~IEEE,}
        Zhilin~Chen,~\IEEEmembership{Member,~IEEE,}
        and~Wei~Yu,~\IEEEmembership{Fellow,~IEEE}% <-this % stops a space
\thanks{The authors are with The Edward S.\ Rogers Sr.\ Department of Electrical and Computer Engineering, University of Toronto, Toronto, ON M5S 3G4, Canada (e-mails:\{fsohrabi,  zchen,  weiyu\}@ece.utoronto.ca). This work is supported by Huawei Technologies Canada and by the Natural Sciences and Engineering Research Council (NSERC) via the Canada Research Chairs program. The source code for this paper is available at: \protect\url{https://github.com/foadsohrabi/DL-ActiveLearning-BeamAlignment}
}
}
\maketitle
%%%%%%%%%%%%%%%%%%%%%%%%%%%%%%%%%%%
% Abstract
%%%%%%%%%%%%%%%%%%%%%%%%%%%%%%%%%%%
\begin{abstract}
This paper proposes a deep learning approach to the adaptive and sequential beamforming design problem for the initial access phase in a mmWave environment with a single-path channel. For a single-user scenario where the problem is equivalent to designing the sequence of sensing beamformers to learn the angle of arrival (AoA) of the dominant path, we propose a novel deep neural network (DNN) that designs the adaptive sensing vectors sequentially based on the available information so far at the base station (BS). By recognizing that the AoA posterior distribution is a sufficient statistic for solving the initial access problem, we use the posterior distribution as the input to the proposed DNN for designing the adaptive sensing strategy. However, computing the posterior distribution can be computationally challenging when the channel fading coefficient is unknown. To address this issue, this paper proposes to use an estimate of the fading coefficient to compute an approximation of the posterior distribution. Further, this paper shows that the proposed DNN can deal with practical beamforming constraints such as the constant modulus constraint. Numerical results demonstrate that compared to the existing adaptive and non-adaptive beamforming schemes, the proposed DNN-based adaptive sensing strategy achieves a significantly better AoA acquisition performance.

\end{abstract}


%%%%%%%%%%%%%%%%%%%%%%%%%%%%%%%%%%%%%%%
%			1. Introduction
%%%%%%%%%%%%%%%%%%%%%%%%%%%%%%%%%%%%%%%
\section{Introduction}
\label{sec:intro}

Millimeter-wave (mmWave) communication is expected to be one of the key enablers of future wireless communication systems \cite{rappaport2013millimeter,haider2014cellular}. Operating at mmWave frequencies allows exploiting the underutilized multi-gigabit bandwidth available at \changeg{the mmWave spectrum}, which can be used to address the ever-increasing demand for higher data rates in future wireless systems. Further, thanks to the shorter wavelength of mmWave signals, we can deploy large-scale antenna arrays in relatively small areas. This leads to the advent of the massive multiple-input multiple-output (MIMO) concept for mmWave communications, in which the transceivers with large-scale antenna arrays form highly directional beamformers in order to combat the poor propagation characteristics of mmWave channels \cite{Molisch2017Survey}. However, constructing such directional beamformers requires an accurate estimate of the channel state information (CSI) that must be obtained in the initial access phase.
The CSI acquisition for massive MIMO systems is a challenging task. This is especially so in practical systems where the number of radio-frequency (RF) chains is typically smaller than the number of antennas, due to cost and power consumption considerations.
 In this paper, we focus on the directional beam alignment problem for massive MIMO systems and illustrate the role of machine learning for the CSI acquisition problem in this context. In particular, we treat the simplest scenario, in which
a base station (BS) with a single RF chain communicates with a single-antenna user in a mmWave environment with a single dominant path channel model. For such a system, the initial alignment problem is equivalent to actively learning the angle of arrival (AoA). The AoA \changeW{obtained at the BS} in the initial access phase can then be used for different purposes, including user localization based on AoA fingerprint \cite{Li_localization2008} and downlink beamforming for time division duplex (TDD) systems \cite{Ng2017}.

The problem of designing the optimal adaptive sensing strategy for AoA acquisition, in general, is quite challenging. To make such an AoA acquisition problem more tractable, the existing state-of-the-art methods in the literature advocate selecting the analog sensing vectors from a pre-designed set of sensing vectors called beamforming \changeg{codebook \cite{alkhateeb2014channel,Zhang2017TCOM,Love2017multires}}, typically \changeg{designed} based on sequential bisection of the AoA range. However, in this paper, we make \changeW{an} observation that codebook-free adaptive beamforming can significantly outperform the state-of-the-art codebook-based adaptive beamforming methods for AoA acquisition in the initial access phase of a mmWave communication. In particular, this paper shows that \changeg{a} deep learning framework can play a crucial role in \changeg{the} effective design of such a codebook-free adaptive beamforming approach. The main reason that deep learning is well-suited for this design is that it can learn how to directly map all the available information at the BS to the sequence of adaptive sensing vectors without restricting them to be selected from a particular pre-designed codebook.

Further, for \changeW{RF-chain-limited systems,} it is desirable from the practical point of view that the designed sensing vectors in the initial access phase \changeWW{are realizable using} simple analog components such as analog phase shifters. Accordingly, in practice, we are interested in designing sensing vectors that satisfy the constant modulus constraint \cite{el2013spatially,sohrabi2016hybrid,Sohrabi2017OFDM}. This paper shows that while the existing codebook-based \changeW{hierarchical beamformer design struggles} to include such practical constraints, the proposed deep learning framework can easily consider the constant modulus constraint. \changeF{Thus, while conventional codebook-based hierarchical beamforming 
would have the advantage of requiring less hardware programmability, the proposed codebook-free beamforming approach has the advantage that it can more readily account for the implementation
constraints of RF-chain limited systems.}

%Prior Work
\subsection{\changeg{Prior Work}}
The early practical works on the mmWave initial access problem consider the exhaustive linear beam search strategies, which scan over all possible beam directions to select the best beam, e.g., \cite{Nitsche2014IEEE}. Such beam searching strategies, which have been proposed in IEEE $802.15$ standards, require a long initial access phase where the initial access time grows linearly with the AoA estimation resolution. The follow-up works show that the delay for AoA acquisition in the initial access phase can be significantly reduced by using carefully designed hashing functions \cite{Abari2016} or by adopting iterative search strategies, e.g., \cite{Giordani2016,alkhateeb2014channel}.
In particular, \cite{Abari2016} proposes a non-adaptive beamforming strategy based on random hash functions to generate sensing vectors and shows that in the \changeFF{high signal-to-noise-ratio (SNR) regime}, the AoA acquisition time grows only logarithmically with the AoA estimation resolution. The paper \cite{alkhateeb2014channel} develops a codebook-based adaptive beamforming approach in which the AoA acquisition time in the high SNR regime also scales logarithmically with the AoA estimation resolution. In doing so, \cite{alkhateeb2014channel} proposes a hierarchical codebook that, in the noiseless setting, allows for an adaptive bisection search over the angular space.

By utilizing the same hierarchical beamforming codebook in \cite{alkhateeb2014channel}, the authors \changeg{of} \cite{Tara2019Active} propose an alternative adaptive beamforming strategy, called \textit{hierarchical posterior matching} (hiePM), which accounts for the measurement noise statistics and selects the beamforming vectors from the hierarchical codebook based on the AoA posterior distribution. It is shown that the hiePM algorithm in \cite{Tara2019Active}, which is mainly devised from the algorithms for sequential noisy search strategies in \cite{Tara2016ITW} and active learning from imperfect labelers in \cite{Tara2016NIPS}, can achieve better performance as compared to the bisection algorithm in \cite{alkhateeb2014channel}.
While the original hiePM algorithm is restricted to the scenario \changeg{in which} the fading coefficient of the single-path channel is known at the BS, recently proposed variants of hiePM extend the results to the more realistic case in which the fading coefficient is unknown, either by using Kalman filter tracking of the fading coefficient in  \cite{Tara2019sequential} or by using the variational Bayesian inference framework in \cite{Akdim2020spawc}.

Nevertheless, the hiePM algorithms in \cite{Tara2019Active,Tara2019sequential,Akdim2020spawc} still employ the hierarchical codebook, and as a result, their overall performances are governed by the quality of this codebook. This paper aims to show that it is possible to design a better adaptive beamforming strategy by employing a codebook-free deep learning approach. In particular, we propose a deep neural network (DNN) that adaptively designs the sensing vectors based on the currently available information at the BS in order to optimize the final AoA acquisition performance.

%Main Contributions
\subsection{\changeg{Main Contributions}}
Similar to the state-of-the-art adaptive beamforming works \cite{alkhateeb2014channel,Tara2019Active,Tara2019sequential,Akdim2020spawc} for the initial access phase, we first consider the AoA detection problem under the simplifying assumption that the AoA is taken from a finite-size grid set. Under this assumption, we formulate the AoA acquisition task as a classification problem in which the sequence of sensing vectors \changeWW{needs to} be designed to minimize the AoA detection error.  \changeb{Motivated by the fact that fully connected neural networks are universal function approximators \cite{hornik1989multilayer}, we propose a DNN architecture to undertake this design. Further, inspired} by the posterior matching methods \cite{Tara2019Active,Tara2019sequential,Akdim2020spawc}, which show that the AoA posterior mass function (PMF) is a sufficient statistic for adaptive sensing, we use the AoA posterior distribution as the \changeW{main} input feature to the proposed DNN. However, as shown in \cite{Tara2019sequential}, \changeg{the exact computation} of the AoA posterior distribution involves \changeWW{several multi-dimensional integrals}; \changeg{this makes} the use of \changeW{the} exact posterior distribution computationally infeasible. To address this issue, this paper proposes first to estimate the fading coefficient to compute an approximation of the AoA \changeFF{posterior distribution}, then to consider the approximated AoA posterior distribution as the input to the proposed DNN. This paper investigates two different estimation strategies for estimating the fading coefficients, the minimum mean squared error (MMSE) estimation and the Kalman filter. Numerical results show that the proposed deep \changeg{learning based} adaptive beamforming approach leads to a lower detection error probability as compared to the existing beamforming strategies for the on-grid AoA detection problem. \changelast{Further, we observe that both 
the MMSE and the Kalman filter approaches to fading coefficient estimation lead to excellent AoA detection performance, but the Kalman filter approach requires lower storage and computational complexity and is thus preferable.}

We further extend the proposed deep learning framework to deal with the more practical \changeWW{gridless} AoA estimation problem with an MMSE objective. In this case, the posterior distribution is no longer in the form of a PMF over a finite-size real vector; instead, it is now in the form of a posterior density function (PDF) over a real interval. Accordingly, we approximate the posterior probabilities that the AoA lies within some small intervals and pass them as inputs to the DNN. Numerical results show that the  proposed DNN-based adaptive beamforming method also outperforms the existing beamforming strategies for the \changeWW{gridless} AoA estimation problem with the MMSE objective.

By recognizing that the practical analog sensing vectors are implemented with simple analog phase shifters, we \changeW{further show} how the proposed DNN architecture can take the constant modulus norm constraint into account. In particular, we propose an appropriate normalization layer that enforces the constant modulus constraint to the sensing matrix elements. Numerical experiments show that while the conventional codebook-based sensing vectors cannot be accurately implemented under constant modulus norm constraint, the proposed DNN with or without the fading information achieves excellent AoA acquisition performance, indicating that the proposed deep learning framework can efficiently deal with the practical constant modulus constraint.

\changeb{We note that deep learning has shown great potential in dealing with complex optimization problems in various wireless communications settings, e.g., for power control in cellular systems \cite{Larsson2020PowerControl}, for MIMO beamforming, \cite{Yang2020,Foad2021FDD,Huang2020FastDL}, and for reconfigurable intelligent surface design \cite{zhang2020deep,Tao2020IRS}.}
More closely to the topic of this paper, the application of deep learning for mmWave beam alignment and beam selection has been adopted in some recent works, e.g., \cite{GeffLi2020MLBA,Rezaie2020ICC}. However, these works only focus on mapping the available information at the BS, in the form of received signals \cite{GeffLi2020MLBA} or location information \cite{Rezaie2020ICC}, to the final downlink beamformer while assuming that the sensing vectors are fixed. To the best of our knowledge, this paper is \changeg{among} the first to address the adaptive sensing design problem for \changeFF{initial beam alignment} by employing machine learning methods and show the advantages of \changeW{optimized sensing} over the conventional codebook-based adaptive beamforming methods. \changeb{We would like to \changeFF{acknow-ledge} a \changelast{similar approach} \cite{abbas2021BA} that also develops a deep learning architecture to tackle the beam alignment problem. 
} 

%Paper Organization and Notations
\subsection{\changeg{Paper Organization and Notations}}
The remainder of this paper is organized as follows. Section~\ref{sec:sys} introduces the system model and the problem formulation for designing an optimal adaptive beamforming strategy for AoA acquisition in a single-path mmWave environment. Section~\ref{sec:motivation} motivates the use of deep learning for solving the initial alignment problem by providing a simple one-stage design example. Section~\ref{sec:ongrid} presents the proposed deep learning framework for designing a codebook-free adaptive beamforming strategy under the assumption that the AoA is taken from a grid set. Section~\ref{sec:ongrid} extends the proposed deep learning framework to deal with the more practical \changeWW{gridless} AoA estimation problem.
Section~\ref{sec:sims} provides the implementation details and simulation results. Finally, conclusions are drawn in Section~\ref{sec:conclusion}.

This paper uses \changeW{lower-case letters for scalar variables}, lower-case bold-face letters for vectors, and upper-case bold-face letters for matrices. The real part and the imaginary part of a complex matrix $\bV$ are respectively given by $\Re(\bV)$ and $\Im(\bV)$. The element-wise absolute value and the element-wise phase of a complex vector $\bv$ are respectively given by $|\bv|$ and $\angle{\bv}$. We use the notations $\odiv$ and $(\cdot)^{\Elemsquare}$ to denote the element-wise division and the element-wise square, respectively. Further, we use the superscripts $(\cdot)^T$, $(\cdot)^H$, $(\cdot)^{-1}$, and $(\cdot)^\dagger$ to denote the transpose, the Hermitian transpose, the inverse, and the pseudoinverse of a matrix, respectively.  The identity matrix with appropriate dimensions is denoted by $\mathbf{I}$; the all-ones vector with appropriate dimension is denoted by $\mathbf{1}$; $\mathbb{R}^{m\times n}$ denotes an $m$ by $n$ dimensional real space; $\mathbb{C}^{m\times n}$ denotes an $m$ by $n$ dimensional complex space; $\mathcal{CN}(\mathbf{0},\mathbf{R})$ and $\mathcal{N}(\mathbf{0},\mathbf{R})$ respectively represent the zero-mean circularly symmetric complex and real Gaussian distributions with covariance matrix $\mathbf{R}$. The notations $\Tr(\cdot)$, $\operatorname{log}_{2}(\cdot)$, $\operatorname{log}_{10}(\cdot)$, and $\mathbb{E} [\cdot] $ represent the trace, binary logarithm, decimal logarithm, and expectation operators, respectively. Finally, \changer{$\|\cdot\|_2$ indicates the Euclidean norm of a vector.}
% and $|\cdot|$ represents the absolute value of a scalar.}

%%%%%%%%%%%%%%%%%%%%%%%%%
%	II) System Model and Problem Formulation
%%%%%%%%%%%%%%%%%%%%%%%%%
\section{System Model and Problem Formulation}
\label{sec:sys}
\changeg{C}onsider a mmWave communications setup in which a BS with $M$ antennas and a single RF chain serves a single-antenna user. To establish a reliable link between the BS and the user, a uplink pilot channel training procedure consisting of $\tau$ time frames \changeg{is used}, where the user transmits uplink pilots $\{x_t\}_{t=1}^{\tau}$, satisfying the power constraint $|x_t|^2\leq P$. Due to the single RF chain constraint, the BS observes the baseband received pilot signals after combining analog signals at the antennas by employing analog beamforming (or sensing) vectors $\{\bw_t\}_{t=1}^{\tau}$, i.e.,
%1
\begin{equation}
\label{eq:meausreModel}
y_t = \bw_t^H \bh x_t + \bw_t^H \bz_t, \quad \forall t \in \{ 1,\ldots,\tau\},
\end{equation}
where $\bh \in \mathbb{C}^N$ is the vector of channel gains between the BS and the user, $\bz_t\sim \mathcal{CN}(\mathbf{0},\bI)$ is the additive white Gaussian noise, and without loss of generality the uplink pilots are set to $x_t = \sqrt{P},\vspace{2pt} \forall t$, and 
the beamforming vectors are assumed to have unit norm \changeg{$\|\bw_t\|^2_2 = 1, \forall t$}. Note that the analog beamformer $\bw_t$ in \changeW{RF-chain-limited systems} is typically implemented via a network of phase shifters, and accordingly, $\bw_t$ needs to further satisfy the constant modulus constraint, i.e., $|w_i^{(t)}| = \tfrac{1}{\sqrt{M}}, \forall i,t$ where $w_i^{(t)}$ is the $i$-th element of $\bw_t$. In this paper, we consider two scenarios \changeg{both with and without the constant modulus constraint.}

We further assume that the mmWave channel between the BS and the user can be modeled by a single dominant path \cite{alkhateeb2014channel}, i.e., 
%2
\begin{equation}
\bh = \alpha \mathbf{a}(\phi),
\end{equation}
where $\alpha \sim \mathcal{CN}(0,1)$ is the complex fading coefficient, $\phi$ is the angle of arrival (AoA), and $\mathbf{a}(\cdot)$ is the array response vector. In a uniform linear array configuration with $M$ antenna elements, the array response vector can be modeled as:
%3
\begin{equation}
\mathbf{a}(\phi) = \left [ 1, e^{j\tfrac{2\pi }{\lambda} d \sin{\phi} },..., e^{j(M-1)\tfrac{2\pi }{\lambda}d \sin{\phi} }  \right]^T,
\end{equation}
where $\lambda$ is the wavelength and $d$ is the antenna spacing.

For many communications applications including localization \cite{Li_localization2008} and downlink beamforming for time division duplex (TDD) systems \cite{Ng2017}, the BS needs to estimate the AoA, $\phi$, from the $\tau$ baseband received signals, i.e., $\{y_t\}_{t=1}^{\tau}$. This procedure, which is called \changeW{\textit{initial beam alignment},} involves the BS sequentially designing the best sensing vectors $\bw_t$ at each time frame $t = 1,2,...,\tau$, possibly in an adaptive manner  in order to optimize the quality of the eventual AoA estimation. This means that the beamforming vector in time frame $t+1$ can be considered as a mapping from the past observations, i.e., the measurements and the beamforming vectors prior to time frame $t+1$, as:  
%4
\begin{equation}
\bw_{t+1} = \widetilde{\mathcal{G}}_t\left(y_{1:{t}},\bw_{1:t} \right), \quad \forall t\in \{ 0,\ldots,\tau-1\},
\label{eq:beamforming_func}
\end{equation}
where $\widetilde{\mathcal{G}}_t : \mathbb{C}^{t}\times \mathbb{C}^{tM} \rightarrow \mathbb{C}^{M}$ is the adaptive beamforming (sensing) strategy that satisfies the beamforming power constraint\footnote{Note that since no prior observation is available at $t=0$, \changeFF{we use an initial
%the same sensing vector 
%$\bw_1$ 
%is used for all channels. Accordingly, in \eqref{eq:beamforming_func}, we have $
$\bw_1 = \mathcal{G}_0(\emptyset,\emptyset)$, where $\emptyset$ denotes that $\mathcal{G}_0(\cdot,\cdot)$ accepts no inputs while it always outputs the same initial vector $\bw_1$.}}. However, instead of designing the beamforming vector based on all past observations, in this paper, we follow the same strategy as in \cite{Tara2019Active}, where $\bw_{t+1}$ is designed based on the AoA posterior distribution at time $t$, which is \changeg{a sufficient statistic} for the AoA estimation problem. Mathematically, if we denote the AoA posterior distribution at time frame $t$ by ${\pi}_t^{(\phi)}$, then the sensing vector for the next measurement can be written as:
%5
\begin{equation}
\bw_{t+1} = {\mathcal{G}}_t \left( {\pi}_t^{(\phi)} \right), 
\end{equation}
where ${\mathcal{G}}_t(\cdot)$ is the function that determines the adaptive beamforming strategy in time frame $t+1$ based on the AoA posterior distribution in time $t$.

The final AoA estimate $\hat{\phi}$ is obtained as a function of the sensing vectors and the baseband received signals in $\tau$ time frames as $\hat{\phi} = \widetilde{\mathcal{F}}\left(y_{1:\tau},\bw_{1:\tau} \right)$, where $\widetilde{\mathcal{F}}: \mathbb{C}^{\tau}\times \mathbb{C}^{\tau M} \rightarrow \mathbb{R}$ is the AoA estimation scheme. Equivalently, we can obtain the final AoA estimation based on the final AoA distribution as:
%6
\begin{equation}
\hat{\phi} = {\mathcal{F}}\left(\pi_\tau^{(\phi)} \right).
\end{equation}

For simplicity of exposition, in this paper, we first consider a simplifying assumption that the prior distribution of AoA for each channel realization is uniform over a grid of $N$ points, i.e., we assume that $\phi \in \{\phi_1,\phi_2,\ldots,\phi_N\}$ where $\phi_i = \phi_\text{min} + \tfrac{i-1}{N-1} (\phi_\text{max} - \phi_\text{min})$. In this case, the AoA posterior distribution $\pi_t^{(\phi)}$ can be represented by an $N$-dimensional vector, i.e., $\boldsymbol{\pi}_t \in \mathbb{R}^N$, where its $i$-th element is defined as:
%7
\begin{equation}
\pi_i^{(t)} = \mathbb{P}\left(\phi=\phi_i ~|~  y_{1:t},\bw_{1:t}  \right). %\quad\quad \forall i =1,\ldots,N.
\end{equation}
Further, in this setup, the task of the BS is to declare one of the candidates in the grid set as $\hat{\phi}$, and the quality of the established link can be determined in terms of the accuracy of the final estimate of $\phi$; see \cite{Tara2019Active}. One way to pursue this program is to formulate it as a detection (classification) problem as follows:
 %8
 \begin{subequations}\label{eq:ongrid_problem}
  \begin{align} 
\displaystyle{\min_{\{\mathcal{G}_t(\cdot)\}_{t=0}^{\tau-1},\hspace{1pt} \mathcal{F}(\cdot)}} &~~\mathbb{P} \left( \hat{\phi} \not= \phi \right)\\
\text{s.t.}~~~~~~ & ~~ \bw_{t+1} = \mathcal{G}_t\left(\boldsymbol{\pi}_t\right), ~~ \forall t \in \{ 0,\ldots,\tau-1\},
\label{eq:BFscheme}\\
~~~~~ &~~ \hat{\phi} = \mathcal{F}\left(\boldsymbol{\pi}_\tau \right),
\end{align}
 \end{subequations}
in which both $\phi$ and $\hat{\phi}$ belong to the pre-described grid set. 

The joint design of the adaptive beamforming strategy and the AoA estimation scheme by directly solving the problem \eqref{eq:ongrid_problem} can be quite challenging. To make such an AoA estimation problem more tractable, the exiting adaptive beamforming schemes in the literature typically design the sensing vectors by selecting them from a pre-designed set of beamformers, called beamforming codebook. The hierarchical beamforming codebook, initially developed in \cite{alkhateeb2014channel}, is an example of such codebooks that has been widely used for the initial alignment problem, e.g., \cite{alkhateeb2014channel,Tara2019Active,Tara2019sequential,Akdim2020spawc}. In this paper, we aim to show that it is possible to design a better adaptive beamforming strategy if we do not restrict ourselves to use such pre-designed codebooks. In particular, we propose a DNN architecture in which the AoA posterior distribution in time frame $t$ is directly mapped to the beamforming vector for obtaining the next measurement, i.e., $\bw_{t+1}$. In this way, the proposed deep \changeg{learning based} adaptive beamforming strategy is not restricted to any particular codebook, and can achieve better performance as compared to the existing algorithms, e.g., \cite{alkhateeb2014channel,Tara2019Active,Tara2019sequential,Akdim2020spawc}, that utilize the hierarchical codebook. 

This paper also considers a more practical setup in which the AoA is \changeW{a continuous variable in $[-\pi,\pi]$. Here} we need to develop a \changeWW{gridless} AoA estimation scheme that operates directly in the continuous domain. In this setting, we consider the MSE \changeFF{as the performance} metric. \changeg{The} AoA estimation problem in the \changeWW{gridless} case can \changeg{then} be written as the following estimation (regression) problem:
%9
 \begin{subequations}\label{eq:gridless_problem}
  \begin{align} 
\displaystyle{\min_{\{\mathcal{G}_t(\cdot)\}_{t=0}^{\tau-1},\hspace{1pt} \mathcal{F}(\cdot)}} &~~\mathbb{E} \left[ \left( \hat{\phi} -  \phi \right)^2 \right]\\
\text{s.t.}~~~~~~ & ~~ \bw_{t+1} = \mathcal{G}_t\left(\pi_t^{(\phi)} \right), ~~ \forall t\in\{0,\ldots,\tau-1\}\\
~~~~~ &~~ \hat{\phi} = \mathcal{F}\left(\pi_\tau^{(\phi)} \right),
\end{align}
 \end{subequations}
where $\hat{\phi},\phi \in [\phi_{\min},\phi_{\max}]$. In this paper, we show that by applying some minor modifications to the proposed neural network architecture for the on-grid AoA detection, we can extend the proposed deep learning framework to tackle the \changeWW{gridless} AoA estimation problem \eqref{eq:gridless_problem}.
 
%%%%%%%%%%%%%%%%%%%%%%%%%
%	III) Motivation for Using Deep Learning 
%%%%%%%%%%%%%%%%%%%%%%%%%
\section{\changeW{Motivation for Using Deep Learning}}
\label{sec:motivation}

This paper aims to show that the deep learning framework can be efficiently used to sequentially design the sensing vectors in the initial beam alignment procedure of a mmWave communication system.
In this section, we first discuss the potential gains of the sequential design of sensing vectors. In the second part of this section, we show the benefits of utilizing the deep learning framework for the initial alignment problem by analyzing  a simpler version of problem \eqref{eq:gridless_problem}, in which \changeW{the sensing vector is designed over one step.}

\subsection{\changeW{Motivation for Sequential Adaptive Sensing}}
\changeW{For general estimation problems, sequential} adaptive design of sensing vectors can lead to better estimation performance compared to non-adaptive schemes. For instance, consider the adaptive compressed sensing problem, where we want to \changeW{\textit{approximately} recover} a $K$-sparse vector \changeW{$\mathbf{x}\in \mathbb{R}^N$} from $\tau$ linear measurements \changeW{of the form $\mathbf{y} = \mathbf{A} \bx + \bz$,} where the sensing vectors, \changeW{i.e., rows of $\mathbf{A}$,} can be adaptively designed. It is shown in \cite{nakos2018improved} for the noiseless setting that the required measurements to accurately recover $\bx$ can be reduced from \changeW{$O\left(K\log \left(\tfrac{N}{K}\right) \right)$} in the non-adaptive case to \changeW{$O\left(K\log \log \left(\tfrac{N}{K}\right) \right)$} in the adaptive sensing design. Further, in the noisy measurement setup, \cite{Haupt2011Adaptive} shows that adaptive compressed sensing enables 
\changeFF{recovery of sparse signals at lower SNRs} as compared to traditional non-adaptive compressed sensing approaches, \changeW{although we remark that there is also evidence in the literature \cite{Candes2013Fund} showing that in a different setting, for the \textit{exact} sparse recovery problem, the difference between adaptive and non-adaptive sensing is minimal. Thus, the benefit of adaptive sensing does depend on the problem setting.}

%%%%%%%%%%%%% Fig. 1
\begin{figure*}
        \centering
        \begin{subfigure}[b]{0.45\textwidth}
            \centering
            \includegraphics[width=0.95\textwidth]{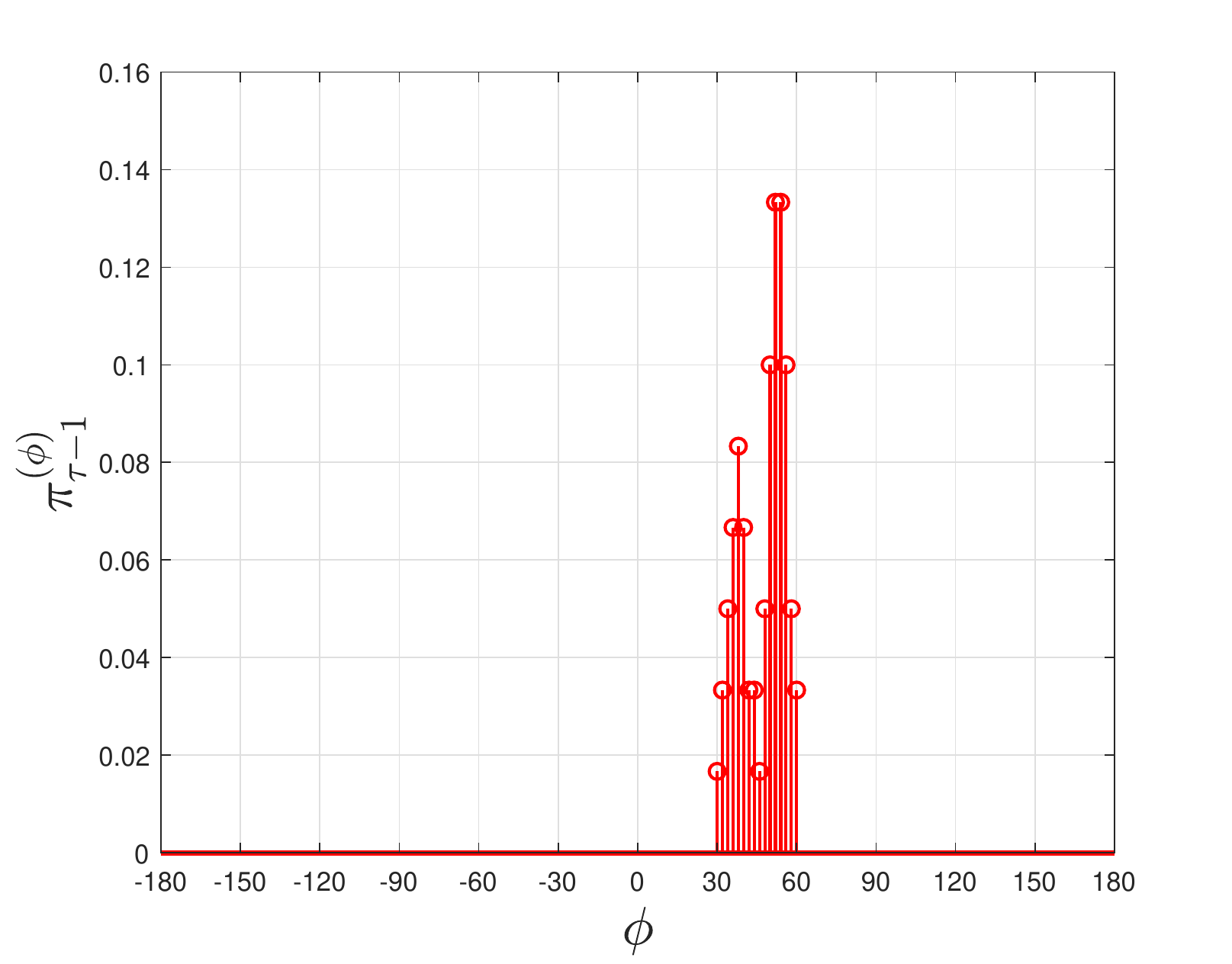}
            \caption[Network2]%
            {{\footnotesize Posterior distribution in time frame $\tau-1$.}}    
            \label{fig:toy_a}
        \end{subfigure}
        \hfill
        \begin{subfigure}[b]{0.45\textwidth}  
            \centering 
            \includegraphics[width=0.7\textwidth]{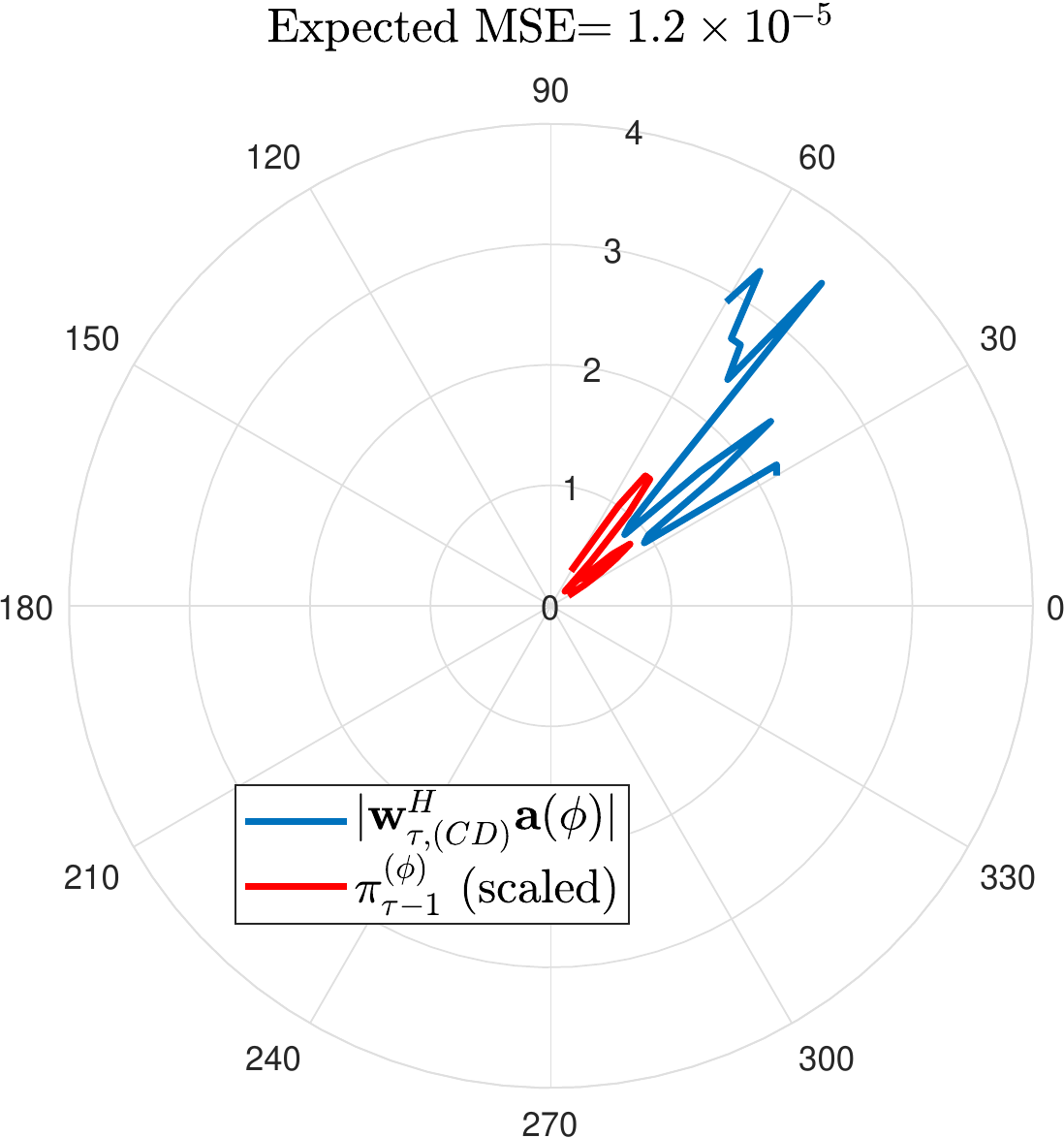}
            \caption[]%
            {{\footnotesize The sensing filter designed by the CD algorithm.}}    
            \label{fig:toy_b}
        \end{subfigure}
        \vskip\baselineskip
        \begin{subfigure}[b]{0.45\textwidth}   
            \centering 
            \includegraphics[width=0.7\textwidth]{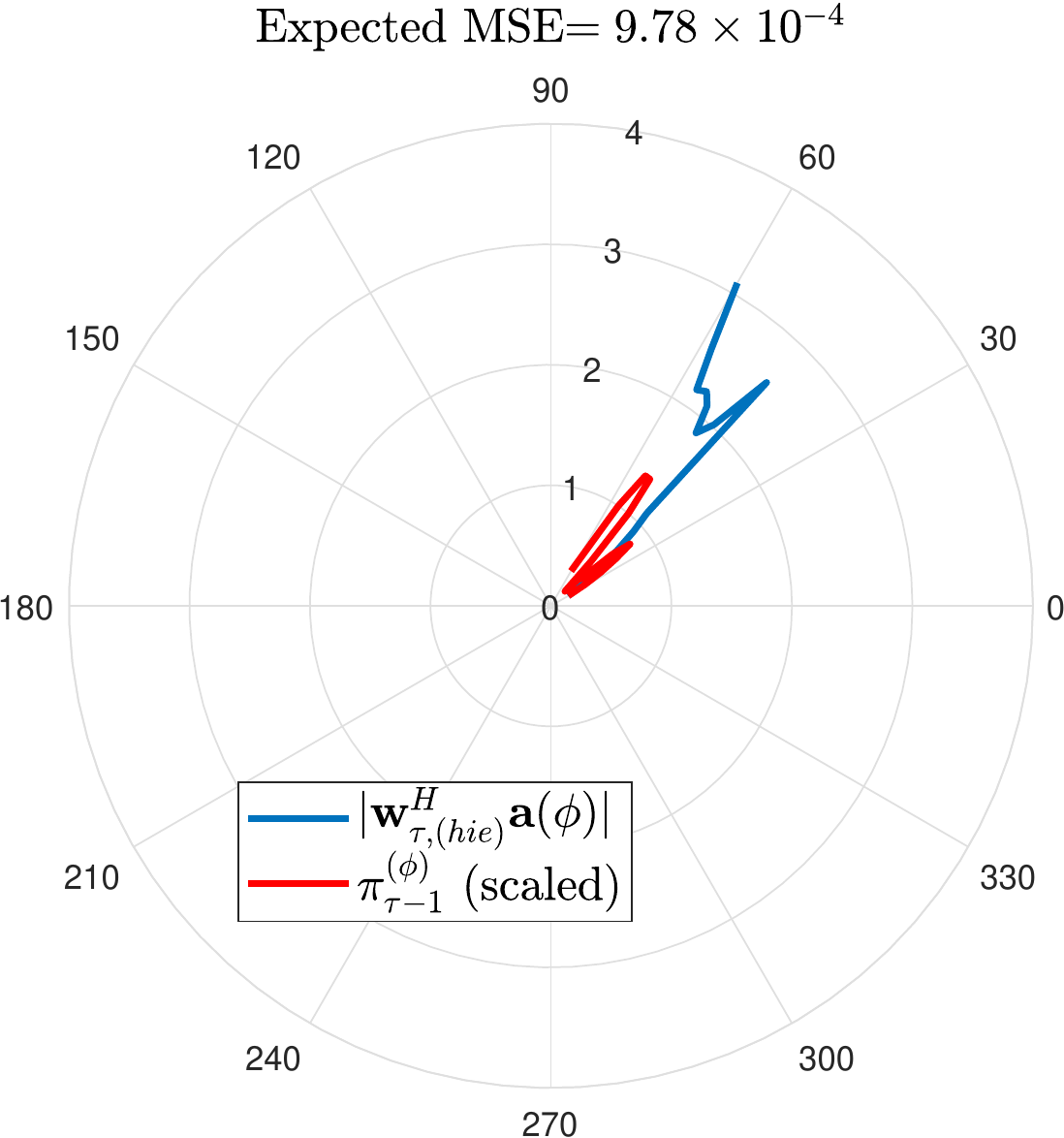}
            \caption[]%
            {{\footnotesize The sensing filter designed by searching over a $4$-stage hierarchical codebook with $30$ sensing vectors.}}    
            \label{fig:toy_c}
        \end{subfigure}
        \hfill
        \begin{subfigure}[b]{0.45\textwidth}   
            \centering 
            \includegraphics[width=0.7\textwidth]{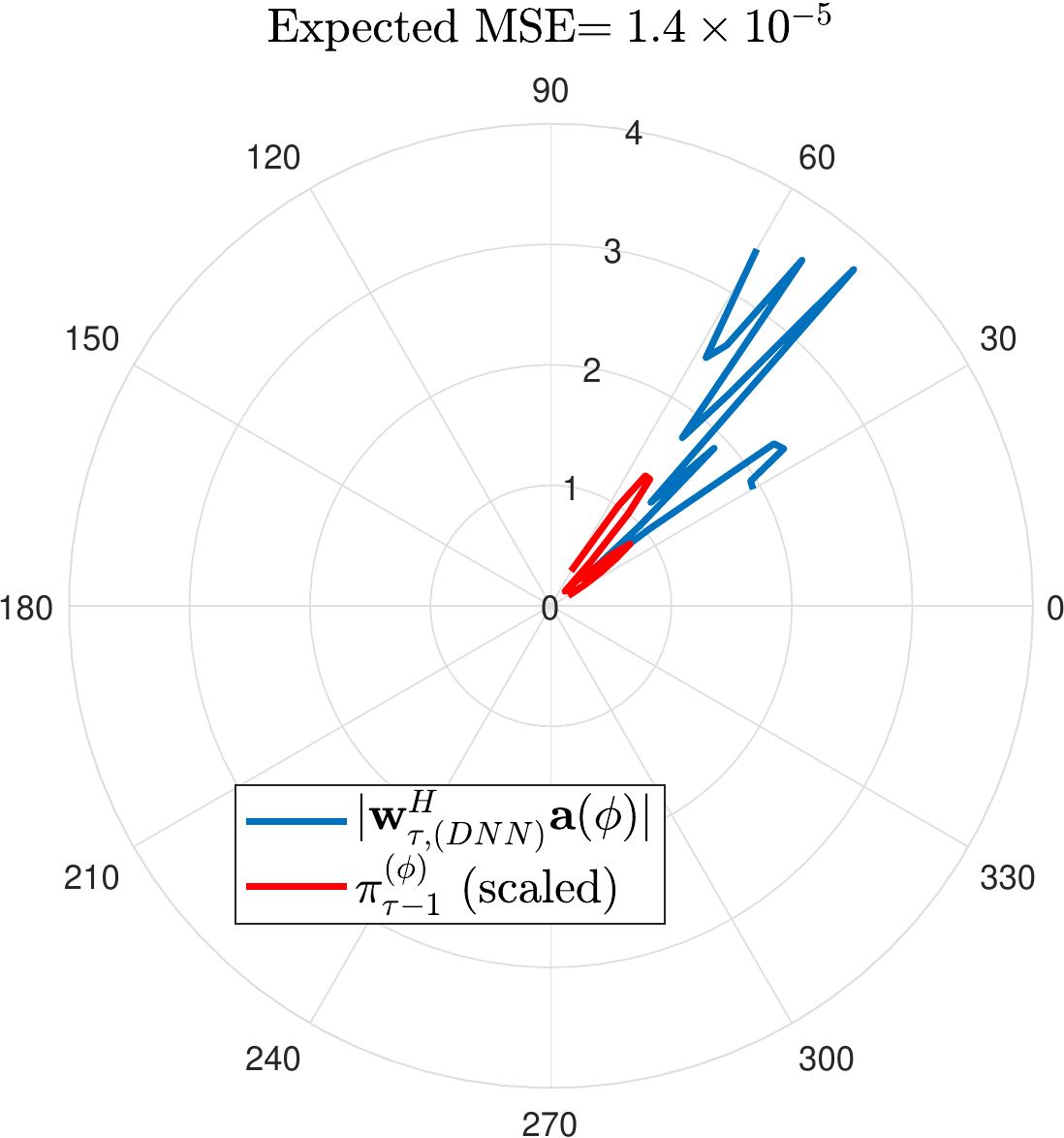}
            \caption[]%
            {{\footnotesize The sensing filter designed by the proposed deep learning framework.}}    
            \label{fig:toy_d}
        \end{subfigure}
        \caption[]{\footnotesize Comparison between different methods for a toy example of designing the sensing vector in time frame $\tau$. } 
\end{figure*}

\changeW{In the adaptive sensing problem for mmWave initial alignment under consideration in this paper}, we deal with a non-linear measurement model as in \eqref{eq:meausreModel}, where we are interested in estimating the AoA parameter $\phi$. Earlier works on the beam alignment problem have already shown the
\changeg{benefits of an adaptive sensing design} for AoA estimation, e.g., \cite{alkhateeb2014channel,Tara2019Active,Tara2019sequential,Akdim2020spawc}, \changeg{which narrows} down the range of the AoA using a hierarchical codebook. The numerical results in this paper confirm \changeg{the performance benefits of more general adaptive strategies}.

\subsection{\changeW{Motivation for Deep Learning}}
\changeW{We now motivate the use of} machine learning for adaptive sensing design by studying the beamforming design problem in \changeW{one step of} the initial alignment procedure. For simplicity, in this part, we assume that the fading coefficient is known, and it is set to $\alpha=1$. Further, we restrict attention to the case of MSE criterion \changeg{without constant modulus constraint}.

%%%%%%%%%%%%% Fig. 2
 \begin{figure*}[t]
 \centering
\includegraphics[width=0.65\textwidth]{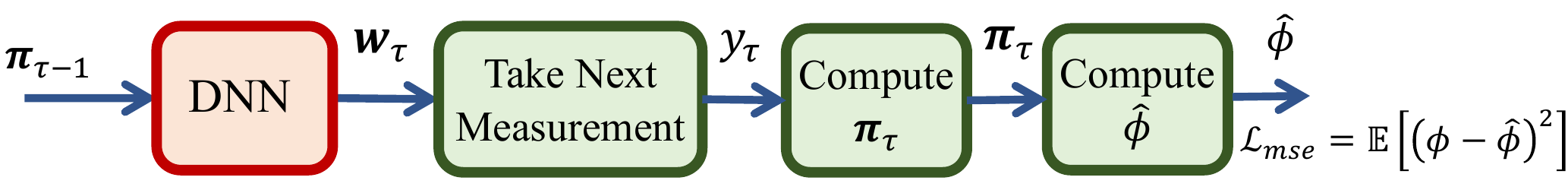}
\caption{The proposed architecture for designing the last sensing vector $\bw_\tau$, given the posterior distribution $\boldsymbol{\pi}_{\tau-1}$.}
\label{fig:DNN_toy}
\end{figure*} 

\changeW{Consider a simple one-step problem of designing} the sensing vector $\bw_\tau$ to minimize the expected MSE of the AoA estimate, given the current posterior distribution $\pi_{\tau-1}^{(\phi)}$. If we fix the sensing vector to $\bw_\tau$ and observe the measurement $y_\tau$, then the \changeW{subsequent} posterior distribution can be computed by applying the \changelast{Bayes'} rule to the measurement model in \eqref{eq:meausreModel} as:
%10
\begin{equation}
{\pi}_{\tau}^{(\phi)} = \frac{{\pi}_{\tau-1}^{(\phi)}e^{-\left|y_{\tau} -  \sqrt{P} \bw_\tau^H \mathbf{a}(\phi)\right|^2}}{\int_{\phi_\text{min}}^{\phi_\text{max}} {\pi}_{\tau-1}^{(\tilde{\phi})} e^{-\left|y_{\tau} -  \sqrt{P} \bw_\tau^H \mathbf{a}(\tilde{\phi})\right|^2} d\tilde{\phi}}.
\end{equation}
Given this AoA posterior distribution, the best MSE estimate for $\phi$ is given by the conditional expectation of $\phi$ given $y_\tau$ as:
%11
\begin{equation}
\label{eq:AoA_estimate_lastStage}
\hat{\phi}({\bw_\tau,y_\tau}) = \mathbb{E} [\phi | y_\tau]  =  \frac{\int_{\phi_\text{min}}^{\phi_\text{max}} \phi ~{\pi}_{\tau-1}^{(\phi)}e^{-\left| y_{\tau} -  \sqrt{P} \bw_\tau^H \mathbf{a}(\phi)\right|^2} d\phi}{\int_{\phi_\text{min}}^{\phi_\text{max}} {\pi}_{\tau-1}^{(\tilde{\phi})} e^{-\left| y_{\tau} -  \sqrt{P} \bw_\tau^H \mathbf{a}(\tilde{\phi}) \right|^2} d\tilde{\phi}}.
\end{equation}
Now, assuming that the true AoA is $\phi_0$, \changer{the squared error (SE)} of the AoA estimate in \eqref{eq:AoA_estimate_lastStage} can be computed as:
%12
\begin{eqnarray}
\label{eq:MSE_formula}
\operatorname{SE}(\bw_\tau,y_\tau,\phi_0) &=& \left(\hat{\phi}({\bw_\tau,y_\tau})  - \phi_0\right)^2.
\end{eqnarray}
The objective function of the beam alignment problem, denoted by $g(\bw_\tau)$, can then be computed by taking the expectation of the SE in \eqref{eq:MSE_formula} over the joint distribution of $y_\tau$ and $\phi_0$ as:
%13
\begin{align} 
g(\bw_\tau) =  \int_{\phi_\text{min}}^{\phi_\text{max}} \int_{-\infty}^{+\infty} &\operatorname{SE}(\bw_\tau,y_\tau,\phi_0) \nonumber\\ &\hspace{-13pt}\times \underbrace{\tfrac{1}{\pi} e^{-\left|y_{\tau} -  \sqrt{P}{{\bw_{\tau}^H}} \mathbf{a}(\phi_0)\right|^2}\boldsymbol{\pi}_{\tau-1}(\phi_0)}_{{\text{joint distribution of $y_\tau$ and $\phi_0$}}} dy_\tau d\phi_0
\end{align}
%\begin{eqnarray}
%g(\bw_\tau) = \int_{\phi_\text{min}}^{\phi_\text{max}} \int_{-\infty}^{+\infty} \operatorname{MSE}(\bw_\tau,y_\tau,\phi_0) . \underbrace{\tfrac{1}{\pi} e^{-\|y_{\tau} -  \sqrt{P}{{\bw_{\tau}^H}} \mathbf{a}(\phi_0)\|^2}\boldsymbol{\pi}_{\tau-1}(\phi_0)}_{{\text{joint distribution of $y_\tau$ and $\phi_0$}}} dy_\tau d\phi_0
%\end{eqnarray}

Without loss of generality, we can formulate the optimization problem of designing the sensing vector $\bw_\tau$ as follows, \changeWW{i.e., with a unit norm constraint on $\bw_\tau$:}
%14
 \begin{subequations}
\label{eq:last_stage_opt}
  \begin{align} 
 \displaystyle{\min_{{\bw_{\tau}}}} & ~~g(\bw_\tau) \\
 \text{s.t.} &~~ \| {{\bw_{\tau}}} \|^2_2 = 1.
 \end{align}
 \end{subequations}

By following the above equations, it can be seen that the objective function of the problem \eqref{eq:last_stage_opt} is a complicated function of the sensing vector. Accordingly, the sensing \changeW{beamformer} design problem for beam alignment, even for a single time frame, is quite complicated. In the rest of this section, we aim to investigate the difficulties of the conventional methods to tackle this challenging problem and subsequently introduce a deep learning framework that can efficiently solve this problem. Toward this aim, we consider a toy example in which the posterior distribution in time frame $\tau-1$, \changeFF{ i.e., $\pi_{\tau-1}^{(\phi)}$,} is given as in \changeW{Fig.~\ref{fig:toy_a}}, \changeWW{where only $16$ AoAs on a grid (of size $16$) have nonzero probabilities, and set $P=10$.} \changeF{We remark that although this section provides a performance comparison between different algorithms only for a particular example of $\pi_{\tau-1}^{(\phi)}$, a similar performance and complexity comparison has been observed for other posterior distributions $\pi_{\tau-1}^{(\phi)}$, including the uniform distribution.}

One conventional method that can be used to tackle the constrained optimization problem in \eqref{eq:last_stage_opt} is the projected gradient descent algorithm. However, the computation of the gradient of the objective function in \eqref{eq:last_stage_opt} with respect to the sensing vector $\bw_\tau$ is not straightforward. Alternatively, we can use a projected coordinate descent (CD) algorithm, in which only one random coordinate of the sensing vector is changed at each iteration, to decrease the objective function at that iteration.  This coordinate update procedure continues until the algorithm converges to a local minimizer of $g(\bw_\tau)$.
By applying such a CD algorithm to the problem \eqref{eq:last_stage_opt} with the current posterior distribution $\pi_{\tau-1}^{(\phi)}$ as in \changeW{Fig.~\ref{fig:toy_a}}, we obtain the design for the sensing vector $\bw_{\tau,(\text{CD})}$. \changeW{Experimentally, this procedure leads to an} expected MSE of $1.2\times 10^{-5}$. The shape of the sensing filter for this design, i.e., $|\bw_{\tau,(\text{CD})}^H \mathbf{a}(\phi)|$, is illustrated in Fig.~\ref{fig:toy_b}. While the expected MSE achieved by the CD algorithm is excellent, it takes several days for this algorithm to converge to the solution on a 4GHz Intel Core(TM) i7-8086K CPU.

The codebook-based beamforming approaches in the literature aim to reduce this computational burden by searching over a predesigned codebook. To investigate the performance of such codebook-based beamforming schemes for the considered problem, we employ a $4$-stage hierarchical codebook containing $30$ beamforming vectors designed according to \cite{alkhateeb2014channel}. The best sensing vector selected from this hierarchical codebook, i.e., $\bw_{\tau,\text{(hie)}}$, achieves the expected MSE of $9.78\times 10^{-4}$, and the filter shape of this design is illustrated in \changeW{Fig.~\ref{fig:toy_c}}.  From this experiment, we observe that although utilizing the hierarchical codebook significantly reduces the computational complexity, its achieved expected MSE is $70$ times higher than the expected MSE of the CD algorithm. \changeg{This} experiment indicates that the codebook-based beamforming schemes fail to reach the optimal design for the initial beam alignment problem.

Finally, we introduce a deep learning framework to design $\bw_\tau$. In particular, we propose to employ an $L$-layer \changeg{fully connected} DNN that takes the non-zero elements of the current posterior distribution ${\pi}_{\tau-1}^{(\phi)}$, denoted by $\boldsymbol{\pi}_{\tau-1}$, and outputs the design for the next sensing vector $\bw_\tau$, i.e., 
%15
\begin{equation} 
\label{eq:dnn_toy}
\widetilde{\bw}_{\tau} = \widetilde{\sigma}_L\left(\widetilde{\bA}_L  \widetilde{\sigma}_{L-1}\left(\cdots \widetilde{\sigma}_1\left(\widetilde{\bA}_1\boldsymbol{\pi}_{\tau-1} +\widetilde{\bb}_1 \right)  \cdots \right) + \widetilde{\bb}_L\right), 
\end{equation}
where $\left\{\widetilde{\bA}_\ell, \widetilde{\bb}_\ell \right\}_{\ell=1}^L$ is the set of the trainable weights and biases of the DNN, $\widetilde{\sigma}_\ell$ is the activation function of the $\ell$-th layer with \changer{$\widetilde{\sigma}_L(\cdot) = \tfrac{\cdot}{\| \cdot \|_2}$}, and $\widetilde{\bw}_{\tau}$ is the real representation\footnote{To use the existing deep learning libraries that only support real-value operations, we consider the beamforming vector's real representation as the output of the DNN.} of $\bw_\tau$, i.e., $\widetilde{\bw}_{\tau} = \left[ \Re\left({\bw}_{\tau}^T\right), \Im{\left( {\bw}_{\tau}^T\right)}\right]^T$.
The output of the DNN can then be used to take the next measurement, \changeg{to} update the posterior distribution, and finally \changeg{to} compute the AoA estimate; see the overall network in Fig.~\ref{fig:DNN_toy}. To train such a network, we can employ \changeg{a stochastic gradient descent (SGD) algorithm} to minimize the expected MSE. In the training phase, we generate a random $\boldsymbol{\pi}_{\tau-1}$ for each mini-batch of size $10^4$, then generate the true AoAs within each mini-batch according to the distribution $\boldsymbol{\pi}_{\tau-1}$. In the test phase, we consider the posterior distribution in \changeW{Fig.~\ref{fig:toy_a}}, and obtain the designed sensing vector $\bw_{\tau,\text{(DNN)}}$ from the output of the trained DNN. From this experiment, we observe that the expected MSE of a trained DNN with $L=4$ layers is $1.4\times 10^{-5}$, which is almost \changeW{identical} to the optimal performance achieved by the CD algorithm. Further, we observe that for \changeg{this} toy example, the proposed DNN can be trained in \changeW{a} few minutes (on a NVIDIA GeForce RTX 2080 GPU), and once the network is trained, the sensing vector design can be obtained in almost real time by following the simple computations in \eqref{eq:dnn_toy}. The other advantage of the proposed deep learning framework is the generalizability of the trained DNN for different \changeFF{posterior distributions} $\boldsymbol{\pi}_{\tau-1}$. In particular,  unlike the conventional methods, which require \changeg{solving} the problem from scratch for each realization of the posterior distribution in time frame $\tau$,  the trained DNN is capable of designing the sensing vector for a class of posterior distributions used in the training phase. These features of the deep learning approach motivate us to employ \changeW{deep learning} for the general beam alignment problem \eqref{eq:gridless_problem}. 
 
%%%%%%%%%%%%%%%%%%%%%%%%%
%	IV) Codebook-Free Adaptive Beamforming for On-Grid AoA Detection
%%%%%%%%%%%%%%%%%%%%%%%%%
\section{Codebook-Free Adaptive Beamforming\\ for On-Grid AoA Detection}
\label{sec:ongrid}

 In this section, we extend the proposed deep learning framework in the previous section to obtain the codebook-free adaptive beamforming strategy for the on-grid AoA detection problem in \eqref{eq:ongrid_problem}, \changeFF{where the AoA is taken from an $N$-point grid set}. As the first step and to fix ideas, we make the simplifying (and unrealistic) assumption that the fading coefficient $\alpha$ is perfectly known at the BS (similar to  \cite{Tara2019Active}). This assumption is removed in Section~\ref{sec:offgrid_unknown}, where we consider a more practical setup in which only the statistical information of $\alpha$ is available. 

%%%%%%%%%%%%%%%%%%%%%%%%%
%	IV.A) Known Fading Coefficient Scenario
%%%%%%%%%%%%%%%%%%%%%%%%%
\subsection{Known Fading Coefficient Scenario}
\label{sec:offgrid_known}
\changeW{We begin by tackling} the AoA detection problem in \eqref{eq:ongrid_problem} for the case that the complex fading coefficient $\alpha$ is assumed to be fully known at the BS. As discussed earlier, under the \changeg{on-grid} assumption, the posterior distribution in time frame $t$ can be represented by an $N$-dimensional vector $\boldsymbol{\pi}_t$. By applying the \changelast{Bayes'} rule to the measurement model in \eqref{eq:meausreModel} under the assumption that the fading coefficient $\alpha$ is known, the posterior distribution in time frame $t+1$ can be computed as follows: 
%16
\begin{equation}
\label{eq:posterior1}
\pi_i^{(t+1)} = \frac{\pi_i^{(t)}f\left(y_{t+1}\big|\phi=\phi_i,\bw_{t+1} = \widetilde{\mathcal{G}}_t \left( \boldsymbol{\pi}_t \right)\right)}{\sum_{j=1}^N \pi_j^{(t)}f\left(y_{t+1}\big|\phi=\phi_j,\bw_{t+1} = \widetilde{\mathcal{G}}_t \left( \boldsymbol{\pi}_t \right)\right)},
\end{equation}
where
\begin{equation}
\label{eq:cond_dist} \nonumber
f\left(y_{t+1}\big|\phi=\phi_i,\bw_{t+1} = \widetilde{\mathcal{G}}_t ( \boldsymbol{\pi}_t) \right) = \tfrac{1}{\pi}e^{-\left|y_{t+1}- \sqrt{P}\alpha\bw_{t+1}^H\mathbf{a}(\phi_i)\right|^2}
\end{equation}
is the conditional distribution of the measurement $y_{t+1}$ given the AoA and the beamforming vector in time frame $t+1$. Once the pilot training phase is completed at the end of the time frame $\tau$, we declare the angle in the grid set with the maximum posterior probability as the final estimate of $\phi$, i.e.,
%17
\begin{equation}
\hat{\phi} =  \phi_{i^\star}, \text{~~~where~} i^\star = {\argmax_{i}}~ \pi_i^{(\tau)}.
 \end{equation}

In the above active AoA detection procedure, the only remaining part to design is the function ${\mathcal{G}}_t(\cdot)$ which maps the AoA posterior distribution to the next sensing vector. By following the proposed idea in Section~\ref{sec:motivation}, we use a data-driven framework to undertake this design. In particular, we propose an $L$-layer \changeg{fully connected} DNN that takes the current posterior distribution together with the other available system parameters as the input, i.e., 
%18
\begin{equation}
\mathbf{v}_t = \left[{\boldsymbol{\pi}_t^T,P,t}\right]^T,
\end{equation}
and outputs the beamforming vector for the next measurement as:
%19
\begin{equation} 
\label{eq:dnn}
\widetilde{\bw}_{t+1} = \sigma_L\left(\bA_L  \sigma_{L-1}\left(\cdots \sigma_1\left(\bA_1\bv_t +\bb_1 \right)  \cdots \right) + \bb_L\right), 
\end{equation}
where $\left\{\bA_\ell, \bb_\ell \right\}_{\ell=1}^L$ is the set of the trainable weights and biases of the DNN, $\sigma_\ell$ is the activation function of the $\ell$-th layer, and $\widetilde{\bw}_{t+1}$ is the real representation of the beamforming vector in time frame $t+1$. 
Further, in order to ensure that the beamforming vector \changeW{satisfies} the \changeW{prescribed} beamforming constraint, we employ an appropriate normalization layer at the last layer of the DNN. In particular, \changeWW{with only a unit norm constraint,} we choose the last layer's activation function as:
%20
\begin{equation}
\sigma_L(\bu) = \frac{\bu}{\| \bu \|_2},  ~\forall \bu \in \mathbb{R}^{2M},
\end{equation}
and for the constant modulus constraint scenario, we set the last activation function as:
%21
\begin{equation}
\sigma_L(\bu) = \tfrac{1}{\sqrt{M}}\left[ \bu_1^T \odiv \left|\bu_1^{\Elemsquare} + \bu_2^{\Elemsquare}\right|, \bu_2^T \odiv \left|\bu_1^{\Elemsquare} + \bu_2^{\Elemsquare}\right| \right]^T,
\label{eq:activation_hyb}
\end{equation}
where  $\bu = \left[\bu_1^T,\bu_2^T\right]^T$ with $\bu_1,\bu_2\in \mathbb{R}^{M}$.

Note that in the proposed DNN architecture, it would be possible to employ different weights and biases for different time frames. However, in this paper, we propose using a common set of DNN weights and biases for all the time frames, and instead, we consider the time frame index as the input to the DNN. The primary motivation for
this consideration is that such a common DNN structure can potentially lead to a more scalable and faster training procedure by reducing \changelast{the number of trainable parameters.}

%%%%%%%%%%%%% Fig. 3
 \begin{figure}[t]
 \centering
\includegraphics[width=0.50\textwidth]{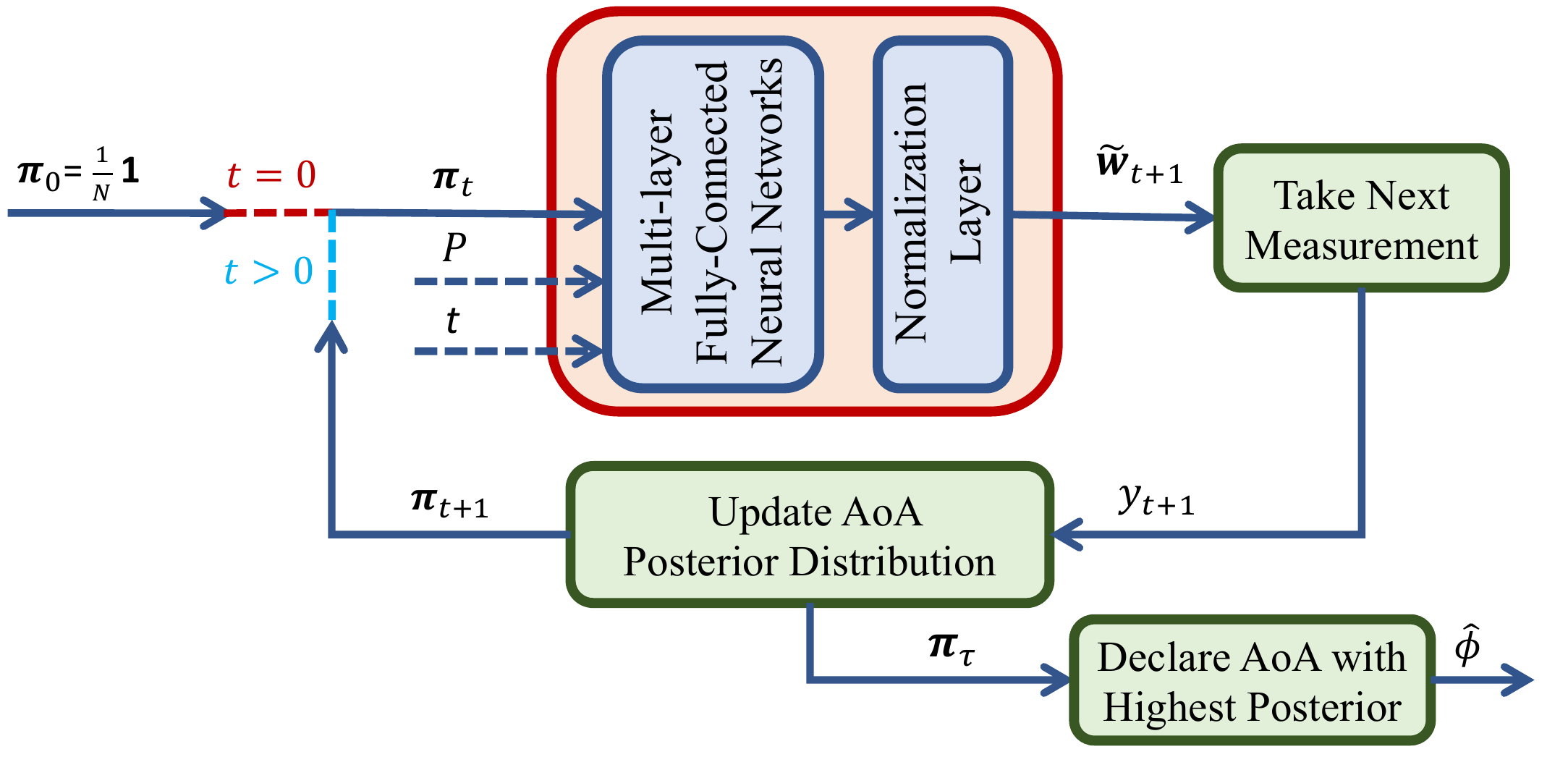}
\caption{The block diagram of the proposed adaptive beamforming strategy for on-grid AoA detection in the initial access phase of a mmWave communications system.}
\label{fig:DNN_feedback1}
\end{figure}

%%%%%%%%% Fig. 4
\begin{figure*}[t]
\centering
 \includegraphics[width=0.75\textwidth]{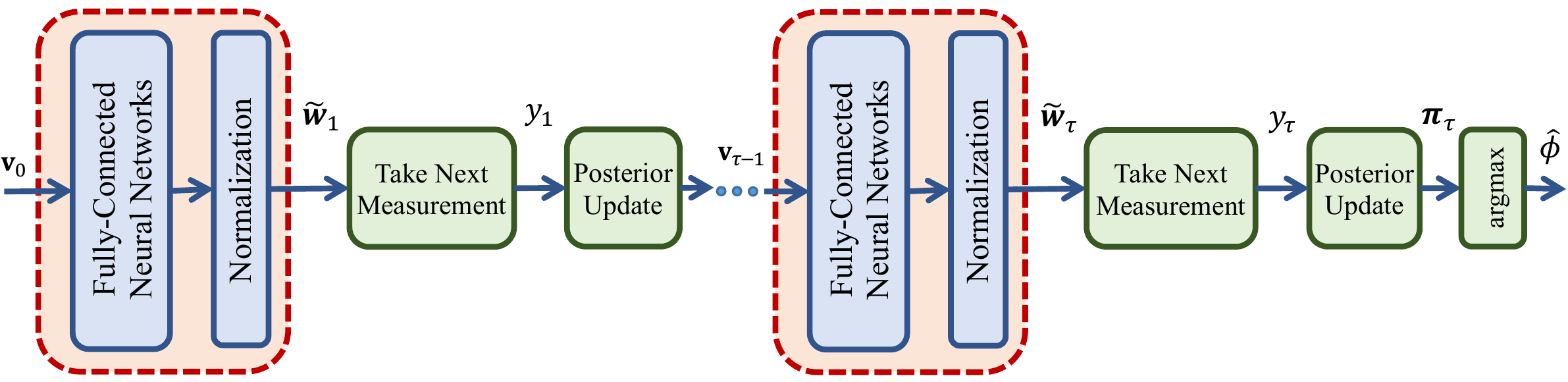}
\caption{The unrolled illustration of the proposed deep \changeg{learning based} AoA detection approach. Note that a common set of DNN weights and biases is employed for obtaining the beamforming vector in different time slots, and the DNN input at each stage consists of the current posterior distribution, the transmitter power, and the time frame index, i.e., $\mathbf{v}_t = \left[{\boldsymbol{\pi}_t^T,P,t}\right]^T$.}
\label{fig:DNN_unrolled}
\end{figure*}

The block diagram of the overall proposed on-grid AoA detection strategy is illustrated in Fig.~\ref{fig:DNN_feedback1}.  As it can be seen from Fig.~\ref{fig:DNN_feedback1}, the adaptive beamforming module implemented by the DNN as in \eqref{eq:dnn} and the posterior distribution update computed by \eqref{eq:posterior1} \changeW{interact} with each other to accurately find the AoA at the end of the time frame $\tau$. Since we assume that there is no prior information available about the AoA distribution, we start the algorithm with a uniform posterior distribution at the time frame $t=0$, i.e., $\boldsymbol{\pi}_0 = \tfrac{1}{N} \mathbf{1}$.  

By unrolling the loop in the proposed adaptive detection algorithm in Fig.~\ref{fig:DNN_feedback1}, we can think of the proposed end-to-end architecture as a very deep neural network architecture shown in Fig.~\ref{fig:DNN_unrolled}. 
The ultimate goal of the DNN network in Fig.~\ref{fig:DNN_unrolled} is to successfully recover the AoA value. This  detection task can be treated as a classification problem for which the categorical cross-entropy is the typical choice of the loss function. Following this observation, we can train the DNN architecture in Fig.~\ref{fig:DNN_unrolled} by employing \changeg{SGD} to minimize the average cross-entropy between
the final posterior distribution at time frame $\tau$ and the one-hot representation of the actual AoA, i.e., $\mathbf{e}_\phi = [e_1^{(\phi)},\ldots,e_N^{(\phi)}]$, where $e^{(\phi)}_i$ is 1 if the actual AoA is $\phi_i$ and $0$ otherwise, i.e., 
%22
\begin{equation}
\mathcal{L} = - \mathbb{E}_\mathbf{q} \left[  \sum_{i=1}^{N} e^{(\phi)}_i \log_2{\pi_i^{(\tau)}} \right],
\end{equation}
where the expectation is over all the stochastic parameters of the system, i.e., $\mathbf{q} \triangleq  [\alpha,\phi,\bz]$ with $\bz = [\bz_1^T,\ldots,\bz_\tau^T]^T$.

\changeF{We remark that the considered problem in this paper is closely related to \emph{reinforcement learning} in the sense that it involves an agent (i.e., BS) who needs to take an action (i.e., design the sensing vector) given the current state (i.e., the current posterior distribution), while the agent's action would affect the next state. However, we do not directly use the conventional tools and concepts in the reinforcement learning literature. Instead, in order to solve the active learning problem, we unroll the beam alignment procedure and consider all $\tau$ beamforming stages together. In this way, we regard the proposed end-to-end architecture as a very deep neural network, with a goal of estimating $\phi$ (see Fig.~\ref{fig:DNN_unrolled}). We then use labeled data to train the DNN in a supervised fashion via the SGD algorithm. In other words, while the nature of the problem is closely related to reinforcement learning, the way that we solve the problem mostly utilizes the principle of supervised learning.}

%%%%%%%%%%%%%%%%%%%%%%%%%%
% IV.B) Unknown Fading Coefficient Scenario
%%%%%%%%%%%%%%%%%%%%%%%%%%
\subsection{Unknown Fading Coefficient Scenario}
\label{sec:offgrid_unknown}
We now deal with a more practical scenario where the fading coefficient $\alpha$ is unknown. Toward this end, we need to modify the expression for updating the posterior distribution in \eqref{eq:posterior1}, which has been derived  for a given $\alpha$ in Section~\ref{sec:offgrid_known}. As shown in \cite{Tara2019sequential}, when $\alpha$ is unknown, the expression of the conditional distribution involves  \changeW{computationally demanding integrals}. Therefore, the exact computation of $\boldsymbol{\pi}_t$ for unknown $\alpha$ may be \changeW{intractable}. 
To address this issue, in this paper, we propose an alternative approach in which the fading coefficient is first estimated in each time frame. Subsequently, the estimate of the fading coefficient is used to compute an approximate of the AoA \changeFF{posterior distribution}. In \changeW{particular}, we investigate two different estimation strategies: i) MMSE estimator, and ii) Kalman Filter.

%%%%%%%%%%%%%%%%%%%%%%%%%%
% Fading Coefficient Estimation via MMSE
%%%%%%%%%%%%%%%%%%%%%%%%%%
\subsubsection{Fading Coefficient Estimation via MMSE}  Assuming that the actual AoA is $\phi_i$, we first seek to estimate the fading coefficient at time frame $t$ from the available measurements, i.e., 
%23
\begin{equation}
\by_t = \mathbf{c}_{i,t} \alpha + \bn_t,
\end{equation}
where $\by_t = [y_1,\ldots,y_t]^T$, $\mathbf{c}_{i,t} = \sqrt{P} \bW_t^H \mathbf{a}(\phi_i)$ with $\bW_t \triangleq [\bw_1,\ldots, \bw_t]$, and $\bn_t \sim \mathcal{CN}(\mathbf{0},\bI)$. Given that the fading coefficient has the standard complex Gaussian distribution, i.e., $\alpha \sim \mathcal{CN}(0,1)$, the best MSE estimate of the fading coefficient at time frame $t$ can be computed as:
%24
\begin{equation}
\label{eq:MMSE}
\hat{\alpha}_i^{(t)} = \left(\mathbf{c}_{i,t}^H \mathbf{c}_{i,t} + 1 \right)^{-1} \mathbf{c}_{i,t}^H \by_t.
\end{equation} 

The MMSE estimate of the fading coefficient in \eqref{eq:MMSE} can then be used to approximate the AoA posterior distribution. In particular, we propose \changeW{to approximate} the AoA posterior distribution by \changeW{regarding} the MMSE estimate for $\alpha$ in \eqref{eq:MMSE} as the actual value of the fading coefficient. With this approximation in place, we can show that the $i$-th element of the AoA posterior distribution at time frame $t+1$ can be computed as:
%25
\begin{equation}
    \pi_i^{(t+1)} = \frac{ \prod_{\tilde{t}=1}^{t+1} e^{-\left|y_{\tilde{t}} -  \sqrt{P} \hat{\alpha}_i^{(t+1)} \mathbf{w}_{\tilde{t}}^H \mathbf{a}(\phi_i) \right|^2}}{\sum\limits_{j =1}^{N}  \prod_{\tilde{t}=1}^{t+1} e^{-\left|y_{\tilde{t}} -  \sqrt{P} \hat{\alpha}_j^{(t+1)} \mathbf{w}_{\tilde{t}}^H \mathbf{a}(\phi_j) \right|^2}}.
    \label{eq:posterior1_mmse}
\end{equation}

\changeWW{Note that the estimate $\hat{\alpha}_i^{(t+1)}$ is based on all the available measurements so far, so it is the most accurate estimate of the fading coefficient. Thus, in \eqref{eq:posterior1_mmse}, we compute  the \changeFF{posterior} in time frame $t+1$ based on $\hat{\alpha}_i^{(t+1)}$.}
Once the approximated \changeFF{posterior distribution is} obtained using \eqref{eq:posterior1_mmse}, analogous to the known fading scenario in Section~\ref{sec:offgrid_known}, we employ a DNN that directly maps the approximated posterior distribution and other available system parameters to the next sensing vector. 
Therefore, the overall proposed AoA detection strategy for the unknown fading coefficient scenario can also be illustrated as in Fig.~\ref{fig:DNN_feedback1}, but in this case the step of updating the \changeFF{posterior distribution} further involves estimating the fading coefficients.

In the simulations, we observe that the proposed MMSE approach for fading estimation leads to an excellent performance. \changeWW{However, this approach requires relatively high memory usage to store all the received signals so far  and the designed sensing vectors, i.e., $\by_t$ and $\bW_t$, in addition to the high computational resources for computing the estimate of the fading component and the posterior distributions, according to \eqref{eq:MMSE} and \eqref{eq:posterior1_mmse}, respectively, in each time frame.} In particular, it can be shown that, in the above MMSE estimation method, the overall complexity of computing the \changeFF{posterior} in time frame $t$ given the fading coefficient estimates is $O(tMN)$, leading to \changeg{a} computational complexity of $O(\tau^2MN)$ for computing the \changeFF{posterior distributions} in all $\tau$ time frames. In the next \changeg{subsection}, we aim to address these challenges of the MMSE estimation method by proposing an alternative estimation algorithm with lower storage and computational complexity based on the Kalman filtering idea. 

%%%%%%%%%%%%%%%%%%%%%%%%
% III.B.2) Fading Coefficient Tracking via Kalman Filter
%%%%%%%%%%%%%%%%%%%%%%%%
\subsubsection{Fading Coefficient Tracking via Kalman Filter} In this approach, the proposed strategy for sequential detection of AoA is \changeW{complemented} by estimating the fading coefficient via a Kalman filter. In this method, we first assume that the conditional probability density of the fading coefficient given the AoA $\phi_i$ and the received signal $\by_t$ is complex Gaussian with mean $\mu_{{\alpha},i}^{(t)}$ and variance $\gamma^{(t)}_{{\alpha},i}$.
 Then, we exploit the \changeg{idea of} Kalman filtering to estimate those means and variances in each time frame. In particular, upon receiving a new observation $y_{t+1}$, the mean and variance of the fading distribution for a presumed AoA $\phi_i$ are updated by Kalman filter according to \cite{kay1993fundamentals}:
 %26
%\begin{subequations}
%\label{eq:Kalman_params}
%\begin{eqnarray}
%\mu_{{\alpha},i}^{(t+1)} &=& \mu_{{\alpha},i}^{(t)} + \frac{ \gamma^{(t)}_{{\alpha},i} {g_{i,t+1}^*} }{ \gamma^{(t)}_{{\alpha},i} |{g_{i,t+1}}|^2 + 1 } (y_{t+1}-\mu_{{\alpha},i}^{(t)}g_{i,t+1} ),\\ 
%\gamma^{(t+1)}_{{\alpha},i} &=& \gamma^{(t)}_{{\alpha},i} \frac{ 1}{ \gamma^{(t)}_{{\alpha},i} |{g_{i,t+1}}|^2 + 1},
%\end{eqnarray}
%\end{subequations}
\begin{subequations}
\label{eq:Kalman_params}
\begin{align}
\mu_{{\alpha},i}^{(t+1)} &= \mu_{{\alpha},i}^{(t)} + \frac{ \gamma^{(t)}_{{\alpha},i} {g_{i,t+1}^*} }{ \gamma^{(t)}_{{\alpha},i} |{g_{i,t+1}}|^2 + 1 } (y_{t+1}-\mu_{{\alpha},i}^{(t)}g_{i,t+1} ),\\ 
\gamma^{(t+1)}_{{\alpha},i} &= \gamma^{(t)}_{{\alpha},i} \frac{ 1}{ \gamma^{(t)}_{{\alpha},i} |{g_{i,t+1}}|^2 + 1},
\end{align}
\end{subequations}
where $g_{i,t+1} = \sqrt{P} ~\mathbf{w}_{t+1}^H \mathbf{a}(\phi_i)$. Considering that the actual fading coefficient \changeg{comes} from a standard complex Gaussian distribution, we initialize the means and variances in time frame $t=0$ as $\mu_{{\alpha},i}^{(0)} = 0$ and $\gamma^{(0)}_{{\alpha},i}=1 ,\forall i$. With the above assumptions in place, the conditional distribution of the measurement $y_{t+1}$ can now be computed as \cite{Tara2019sequential}:
%27
\begin{equation}
f\left(y_{t+1}\big|\phi=\phi_i,\bw_{t+1} = \widetilde{\mathcal{G}}_t ( \boldsymbol{\pi}_t) \right) = \tfrac{1}{\pi} e^{\tfrac{-|y_{t+1}- \mu_{{\alpha},i}^{(t+1)} g_{i,t}|^2}{\gamma^{(t+1)}_{{\alpha},i} |g_{i,{t+1}}|^2 +1}}.
\end{equation}
Finally, we can use the \changelast{Bayes'} rule in \eqref{eq:posterior1} in order to compute the AoA posterior distributions in each time frame as:
%28
\begin{equation}
\label{eq:posterior2}
\pi_i^{(t+1)} = \frac{\pi_i^{(t)} ~ e^{\tfrac{-|y_{t+1}- \mu_{{\alpha},i}^{(t+1)} g_{i,t}|^2}{\gamma^{(t+1)}_{{\alpha},i} |g_{i,{t+1}}|^2 +1}}}{\sum_{j=1}^N \pi_j^{(t)} e^{\tfrac{-|y_{t+1}- \mu_{{\alpha},j}^{(t+1)} g_{j,t}|^2}{\gamma^{(t+1)}_{{\alpha},j} |g_{j,{t+1}}|^2 +1}}}.
\end{equation}

From the above equations, we can see that the parameters of the fading coefficient in \eqref{eq:Kalman_params} \changeg{and} the posterior distribution in \eqref{eq:posterior2} are updated sequentially without the need for storing the entire \changeg{set of} past sensing vectors $\bW_t$, and the corresponding  received measurement $\by_t$. Further, it can be shown that the complexity of computing the \changeFF{posterior} in each time frame is $O(MN)$, leading to the computational complexity of $O(\tau MN)$ for computing the \changeFF{posterior distributions} in all $\tau$ time frames. This means that the posterior update phase in the Kalman filtering method is computationally more efficient than the MMSE estimation method, which has a complexity of $O(\tau^2 MN)$. Such storage and computational benefits provided by the Kalman filter method make it a promising candidate, especially when integrated with the deep learning framework.
%%%%%%%%%%%%%%%%%%%%%%%%%
%	V) Codebook-Free Adaptive Beamforming for On-Grid AoA Detection
%%%%%%%%%%%%%%%%%%%%%%%%%
\section{Codebook-Free Adaptive Beamforming\\ for \changeWW{Gridless} AoA Estimation}
In the previous section, we \changelast{make} the simplifying assumption that the AoA belongs to a grid set of size $N$, and accordingly, \changeg{develop} a deep active AoA learning scheme, which utilizes the $N$-dimensional AoA posterior distribution to design the sequence of sensing vectors. 
This section aims to further extend the proposed active AoA learning scheme to a more realistic setup, where the AoA can be any real number in the range $[\phi_\text{min},\phi_\text{max}]$. Similar to Section~\ref{sec:ongrid}, as the first step to fix ideas, we begin \changeW{with} the simplified scenario in which the fading coefficient is known at the BS.

When the AoA is a continuous \changeFF{variable,} the posterior distribution can no longer be represented with an $N$-dimensional PMF. Instead, the posterior distribution is now in the form of a PDF whose value at any given $\phi \in [\phi_\text{min},\phi_\text{max}]$ is potentially non-zero. In principle, in this scenario, we need to consider this posterior PDF as the input to the DNN. However, since an accurate representation of a generic PDF requires infinitely many samples, it may not be feasible to feed the AoA \changeFF{posterior distribution} to the DNN directly. To overcome this issue, instead of passing the AoA distribution to the DNN, we propose to use the posterior probabilities \changelast{that} the AoA lies within \changelast{finite} intervals. In particular, the entire range of valid AoAs \changelast{is divided} into $N_c$ intervals, where the $i$-th interval is denoted by: 
%29
\begin{equation}
\varphi^i=\left\{\phi ~|~ \phi \in [\phi_\text{min}^i,\phi_\text{max}^i]\right \},
\end{equation}
 with 
%30
\begin{subequations}
\begin{align}
\phi_\text{min}^i &= \phi_\text{min} + (i-1)\Delta\phi_c,\\
\phi_\text{max}^i &= \phi_\text{min} + i ~\Delta\phi_c,\\
\Delta\phi_c &= \frac{\phi_\text{max}-\phi_\text{min}}{N_c}.
\end{align}
\end{subequations}

\begin{figure}[t]
        \centering
        \begin{subfigure}[b]{0.5\textwidth}
            \centering
            \includegraphics[width=\textwidth]{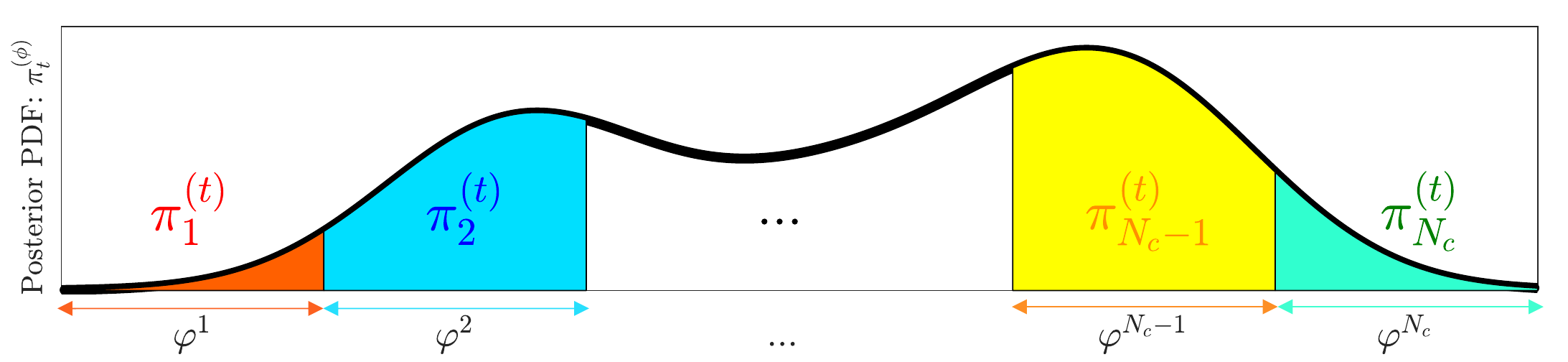}
            \caption[Network2]%
            {{\footnotesize The posterior probabilities by which the AoA lies within intervals $\{\varphi^i\}_{i=1}^{N_c}$ are considered as DNN's inputs.}}    
            \label{}
        \end{subfigure}
        \vskip\baselineskip
        \begin{subfigure}[b]{0.5\textwidth}   
            \centering 
            \includegraphics[width=\textwidth]{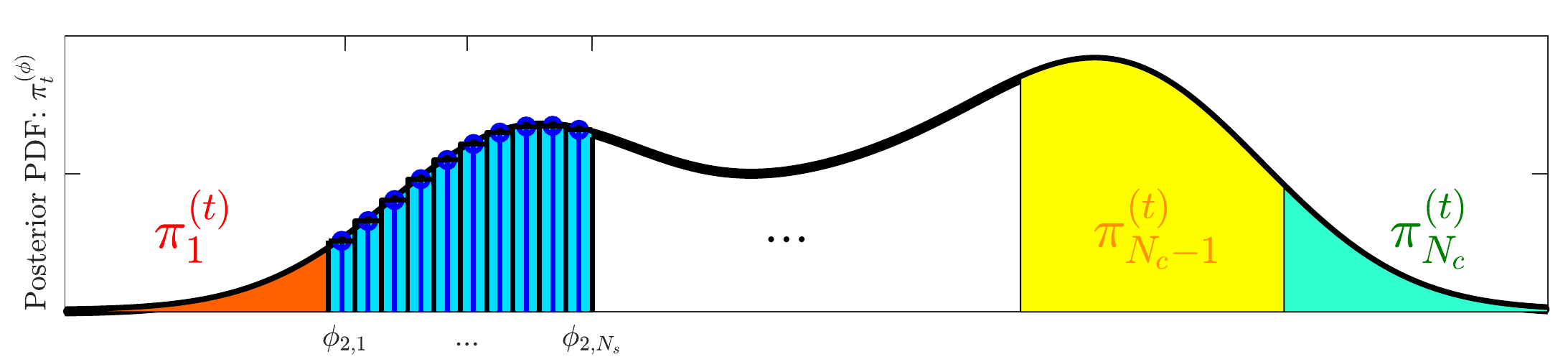}
            \caption[]%
            {{\footnotesize To approximate those posterior probabilities, the midpoint rule approximation is employed.}}    
            \label{}
        \end{subfigure}
        \caption[]{ The proposed procedure to construct the part of the DNN input that depends on the AoA posterior distribution.} 
        \label{fig:dist_sampling}
\end{figure}

Now, the posterior probability (in time frame $t$) that the true AoA $\phi$ belongs to the interval $\varphi^i$ can be computed as:
%31
\begin{equation}
\label{eq:post_prob}
\pi_i^{(t)} = \mathbb{P} \left(\phi \in \varphi^i ~|~ t\right) = \int_{\varphi^i} \pi_{t}^{(\phi)} d\phi, ~~~\forall i \in \{1,\ldots,\changeb{N_c}\},
\end{equation} 
where $\pi_{t}^{(\phi)}$ is the AoA posterior PDF in time frame $t$. By stacking all $N_c$ posterior interval probabilities in \eqref{eq:post_prob}, we can form the posterior probability vector, i.e., $\boldsymbol{\pi}_t = \left[\pi_1^{(t)},\ldots,\pi_{N_c}^{(t)}\right]^T$, that can be considered as the input to the DNN to design the sensing vector in time $t+1$. However, having an integral operation in \eqref{eq:post_prob} within the DNN framework makes the back-propagation step in the SGD-based training methods complicated and computationally expensive. To deal with this issue, we \changeg{approximate} the integral in \eqref{eq:post_prob} using the midpoint rule as follows:
%32
\begin{equation}
\pi_i^{(t)} \approx \sum_{j=1}^{N_{s}} \pi_{i,j}^{(t)} ,
\label{eq:approx_post_prob}
\end{equation}
where for a known $\alpha$ we have:
%33
%\begin{subequations}
\begin{align}
 \pi_{i,j}^{(t)} \displaystyle{\triangleq} \pi_{t}^{(\phi_{i,j})}\Delta\phi_s &= \frac{  \pi_{t-1}^{(\phi_{i,j})}e^{-\left|y_{t}- \sqrt{P}\alpha\bw_{t}^H\mathbf{a}(\phi_{i,j})\right|^2}}{\int_{\phi_\text{min}}^{\phi_\text{max}}  \pi_{t-1}^{(\phi)} e^{-\left|y_{t}- \sqrt{P}\alpha\bw_{t}^H\mathbf{a}(\phi)\right|^2} d\phi} \Delta\phi_s \nonumber\\
&\hspace{-3pt} \stackrel{(a)}{{\approx}} \frac{ \pi_{t-1}^{(\phi_{i,j})}e^{-\left|y_{t}- \sqrt{P}\alpha\bw_{t}^H\mathbf{a}(\phi_{i,j})\right|^2} \Delta\phi_s}{\displaystyle{\sum_{\bar{i}=1}^{N_c} \sum_{\bar{j}=1}^{N_s}} \pi_{t-1}^{(\phi_{\bar{i},\bar{j}})} e^{-\left|y_{t}- \sqrt{P}\alpha\bw_{t}^H\mathbf{a}(\phi_{\bar{i},\bar{j}})\right|^2} \Delta\phi_s} \nonumber\\
&\hspace{-3pt}\stackrel{(b)}{=} \frac{ \pi_{i,j}^{(t-1)} e^{-\left|y_{t}- \sqrt{P}\alpha\bw_{t}^H\mathbf{a}(\phi_{i,j})\right|^2}}{\displaystyle{\sum_{\bar{i}=1}^{N_c} \sum_{\bar{j}=1}^{N_s}} \pi_{\bar{i},\bar{j}}^{(t-1)} e^{-\left|y_{t}- \sqrt{P}\alpha\bw_{t}^H\mathbf{a}(\phi_{\bar{i},\bar{j}})\right|^2}},
\label{eq:posterior_update_equation}
\end{align}
%\end{subequations}
where $(a)$ is due to the midpoint approximation rule in which $\Delta\phi_s = \tfrac{\phi_\text{max}^i -\phi_\text{min}^i}{N_s} = \tfrac{\phi_\text{max} -\phi_\text{min}}{N_c N_s}$, $\phi_{i,j} = \phi_\text{min}^i + \tfrac{2j-1}{2} \Delta\phi_s$, and $N_s$ is the number of considered midpoints, and $(b)$ is due to the definition of $\pi_{i,j}^{(t-1)}$.
An example of the above procedure to construct the part of the DNN input that depends on the AoA posterior distribution is illustrated in Fig.~\ref{fig:dist_sampling}.

%%%%%%%%%%%%% Fig. 5
%%33
%\begin{subequations}
%\begin{align}
% \pi_{i,j}^{(t)} \displaystyle{\triangleq} \pi_{t}^{(\phi_{i,j})}\Delta\phi_s &= \frac{  \pi_{t-1}^{(\phi_{i,j})}e^{-\|y_{t}- \sqrt{P}\alpha\bw_{t}^H\mathbf{a}(\phi_{i,j})\|^2}}{\int_{\phi_\text{min}}^{\phi_\text{max}}  \pi_{t-1}^{(\phi)} e^{-\|y_{t}- \sqrt{P}\alpha\bw_{t}^H\mathbf{a}(\phi)\|^2} d\phi} \Delta\phi_s\\
%&\hspace{-17pt}\stackrel{\text{midpoint rule}}{\approx} \frac{ \pi_{t-1}^{(\phi_{i,j})}e^{-\|y_{t}- \sqrt{P}\alpha\bw_{t}^H\mathbf{a}(\phi_{i,j})\|^2} \Delta\phi_s}{\sum_{\bar{i}=1}^{N_c} \sum_{\bar{j}=1}^{N_s} \pi_{t-1}^{(\phi_{\bar{i},\bar{j}})} e^{-\|y_{t}- \sqrt{P}\alpha\bw_{t}^H\mathbf{a}(\phi_{\bar{i},\bar{j}})\|^2} \Delta\phi_s}\\
%&\hspace{-14pt}\stackrel{\text{def. }\pi_{i,j}^{(t-1)}}{=} \frac{ \pi_{i,j}^{(t-1)} e^{-\|y_{t}- \sqrt{P}\alpha\bw_{t}^H\mathbf{a}(\phi_{i,j})\|^2}}{\sum_{\bar{i}=1}^{N_c} \sum_{\bar{j}=1}^{N_s} \pi_{\bar{i},\bar{j}}^{(t-1)} e^{-\|y_{t}- \sqrt{P}\alpha\bw_{t}^H\mathbf{a}(\phi_{\bar{i},\bar{j}})\|^2}},
%\end{align}
%\end{subequations}
%where $\Delta\phi_s = \tfrac{\phi_\text{max}^i -\phi_\text{min}^i}{N_s} = \tfrac{\phi_\text{max} -\phi_\text{min}}{N_c N_s}$, $\phi_{i,j} = \phi_\text{min}^i + \tfrac{2j-1}{2} \Delta\phi_s$, and $N_s$ is the number of considered midpoints to approximate the integral in \eqref{eq:post_prob}. 

We note that the choices for the number of AoA intervals $N_c$ and the number of samples within each interval $N_s$ in the above procedure provide an accuracy-complexity trade-off, i.e., larger values for $N_c$ and $N_s$ render the approximation in \eqref{eq:approx_post_prob} more accurate, but at the computational complexity cost. We emphasize that when the posterior distribution is more skewed, we need to sample more heavily, i.e., choose higher $N_c$ and $N_s$, to capture such a distribution. 
The skewed AoA distribution phenomenon typically happens in the high SNR regime, where after sufficient \changelast{number of} time frames, the posterior distribution is almost zero \changelast{everywhere} except at around the actual AoA. However, for a practical mmWave communication in which the SNR is typically low or moderate, we numerically observe that a relatively small sampling rate, e.g., $N_c=128$ and $N_s=20$, is sufficient to capture the AoA distribution with the proposed approximation technique.

The other element of the proposed AoA detection strategy in Fig.~\ref{fig:DNN_feedback1} that needs to be modified for the \changeWW{gridless} AoA estimation problem is the last stage of the algorithm \changeW{that provides} the final AoA estimation $\hat{\phi}$. Since we consider the MMSE as the metric for the quality of the AoA estimation, i.e., $\mathcal{L} = \mathbb{E}\left[\left( \phi-\hat{\phi} \right)^2 \right]$, the optimal solution is given by the conditional expectation:
%34
\begin{equation}
\hat{\phi} = \int_{\phi_\text{min}}^{\phi_\text{max}} \phi \hspace{1pt}\pi_{\tau}^{(\phi)}\hspace{1pt} d\phi.
\end{equation}
 Analogous to the previous part, we can approximate this conditional expectation by using the midpoint rule as:
 %35
\begin{equation}
\hat{\phi} \approx \sum_{i=1}^{N_c} \sum_{j=1}^{N_s} \phi_{i,j} \pi_{\tau}^{(\phi_{i,j})}\Delta\phi_s = \sum_{i=1}^{N_c} \sum_{j=1}^{N_s} \phi_{i,j}  \pi_{i,j}^{(\tau)}.
\label{eq:phi_hat_express}
\end{equation}
This means that the AoA MMSE solution can be approximated with the linear combination of the posterior probabilities computed at $N_c\times N_s$ points. In this paper, instead of fixing the weights of this linear combination as in \eqref{eq:phi_hat_express}, we use a \changeg{fully connected} layer with \changeW{trainable} linear activation function that maps $\pi_{i,j}^{(\tau)}$'s to the final AoA estimation.

%%%%%%%%%%%%% Fig. 6
 \begin{figure}[t]
 \centering
\includegraphics[width=0.49\textwidth]{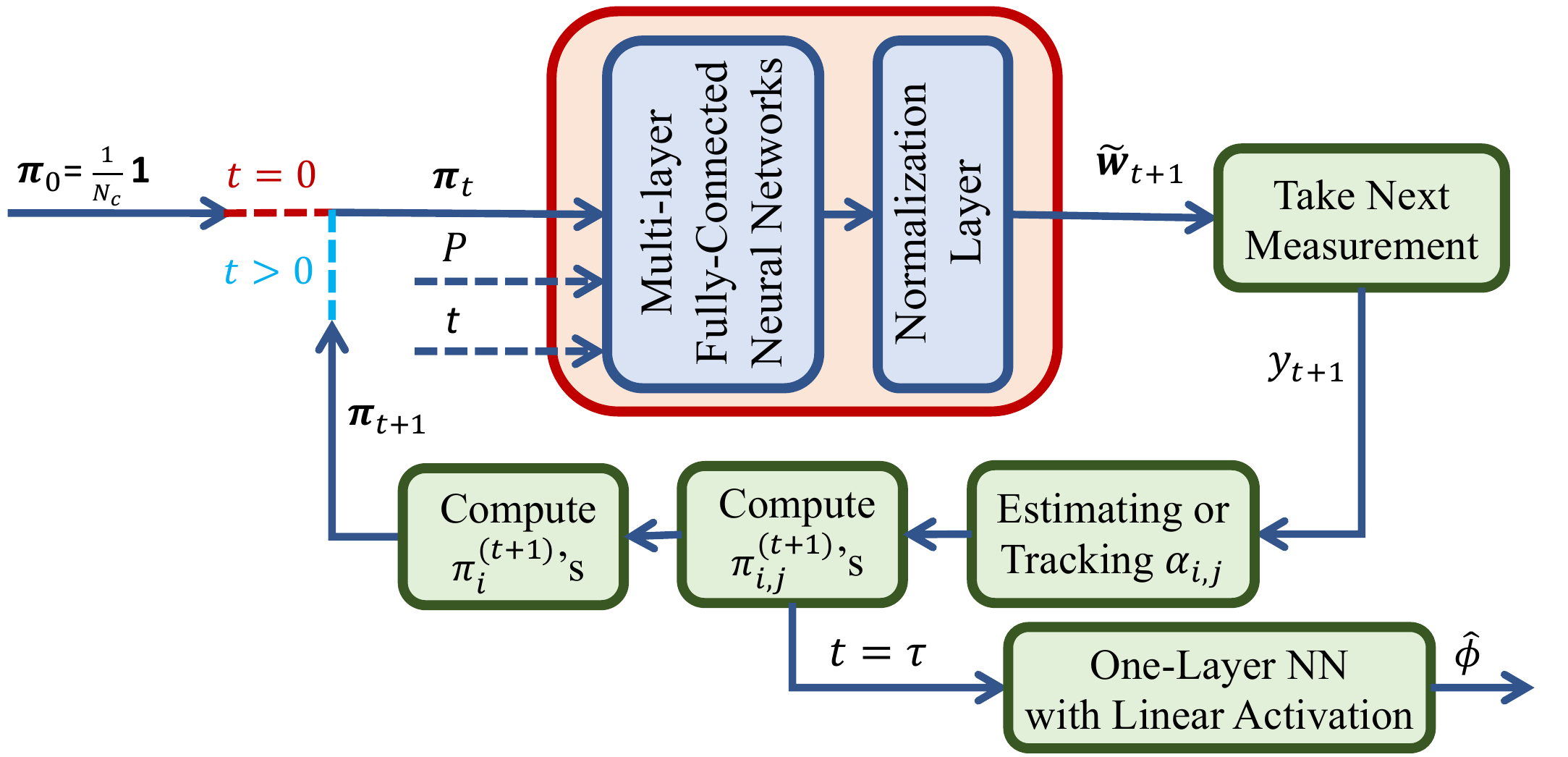}
\caption{The block diagram of the proposed adaptive beamforming strategy for \changeWW{gridless} AoA estimation in the initial access phase of a mmWave communications system.}
\label{fig:DNN_feedback_estimation}
\end{figure}

Finally, we remark that we can easily extend the proposed AoA estimation framework \changeg{to the scenario where} the fading coefficient $\alpha$ is unknown. To do so, we can employ the proposed fading coefficient estimation methods in Section~\ref{sec:offgrid_unknown}. By incorporating such fading estimation techniques,  the overall proposed \changeWW{gridless} AoA estimation can be illustrated as in Fig.~\ref{fig:DNN_feedback_estimation}.

%\section{Adaptive Beamforming under Hybrid Structure Constraint}
%\label{sec:analog}
%
%Finally, this section considers the adaptive beamforming design for the initial access of a mmWave system when the BS employs the hybrid beamforming architecture. So far, we assume that the analog sensing vectors only satisfy the power constraint, i.e., $\|\bw_t\| = 1$. However, the analog beamformer $\bw_t$ in RF-chain-limited systems is typically implemented via a network of phase shifters, and accordingly, $\bw_t$ satisfies the constant modulus constraint, i.e., $|w_i^{(t)}| = \tfrac{1}{\sqrt{M}}, \forall i,t$ where $w_i^{(t)}$ is the $i$-th element of $\bw_t$.
%
%To enforce the constant modulus constraint in the hybrid structure, we adopt the idea in \cite{Lin2020}, where the DNN first outputs the phase of the analog sensing vectors as:
%\begin{equation}
%{\boldsymbol{\Theta}}_{t+1} = \sigma_L\left(\bA_L  \sigma_{L-1}\left(\cdots \sigma_1\left(\bA_1\bv_t +\bb_1 \right)  \cdots \right) + \bb_L\right), 
%\end{equation}
%Subsequently, we apply the entry-wise transformation  $w_i^{(t+1)} = e^{\imath \Theta_{i}^{(t+1)}}$, where $\Theta_{i}^{(t+1)}$ denotes the $i$-th element of the vector 
%${\boldsymbol{\Theta}}_{t+1}$. Such a transformation layer can be implemented using the so-called \textit{Lambda layer} in existing deep learning libraries.  

%%%%%%%%%%%%%%%%%%%%%%%%%%%
%   VI) Numerical Results
%%%%%%%%%%%%%%%%%%%%%%%%%%%
\section{Numerical Results}
\label{sec:sims}

This section illustrates the performance of the proposed deep \changeg{learning based} adaptive beamforming method for initial beam alignment in a mmWave environment. We compare the performance of the proposed method against several existing schemes in the literature. Before presenting the numerical results, we first briefly explain each of the considered AoA detection baselines when the AoA is assumed to be taken from an $N$-point grid set. 

\textit{1) Compressive sensing (CS) with fixed beamforming:} In this approach, we randomly generate the sensing vectors for all $\tau$ frames satisfying the power constraint. Denoting the collection of all sensing vectors by $\bW =  [\bw_1,\ldots,\bw_\tau] $ and the collection of response vectors for all $N$ possible AoAs by $\bA_\text{BS} =  [\mathbf{a}(\phi_1), \ldots, \mathbf{a}(\phi_N)] $, the received signal at the BS in $\tau$ time frames can be written as  $\by_\tau = \bW^H \bA_\text{BS} \bx + \bn_\tau$, where $\bx$ is an unknown $1$-sparse vector. The AoA detection problem can now be cast as finding the support of $\bx$. This sparse recovery problem can be tackled by employing CS  techniques. Here, we adopt a widely-used CS algorithm called orthogonal matching pursuit (OMP) \cite{Tropp2007OMP}. 

\textit{2) Hierarchical codebook with bisection search (hieBS) \cite{alkhateeb2014channel}:} This adaptive beamforming method employs a hierarchical codebook with $S=\log_2(N)$ levels of beam patterns such that each level $s$ consists of a set of 
$2^s$ sensing vectors which partition the AoA search space, i.e., $[\phi_\text{min},\phi_\text{max}]$, into $2^s$ sectors. The beamformer \changeg{for} each sector is designed such that the beamforming gain is almost constant for AoAs within that probing sector, and nearly zero otherwise. In particular, the $k$-th sensing vector in stage $s$ of the hierarchical codebook is obtained as:
%36
\begin{equation}
\bw_{s,k} = \frac{\bA_\text{BS}^\dagger \bg_{s,k}}{\|\bA_\text{BS}^\dagger \bg_{s,k}\|_2},
\label{eq:hie_total_power}
\end{equation}
where $\bg_{s,k}$ is an $N$-dimension vector containing $1$'s in the entries $\left\{(k-1)\tfrac{N}{2^s}+1,\ldots, k\tfrac{N}{2^s}\right\}$, and zeros in the other entries. Further, the sensing vector in each time frame is selected from such a hierarchical codebook using a binary search algorithm, which requires $\tau = 2S = 2\log_2(N)$ time frames for AoA detection from an $N$-point grid set.

\textit{3) Hierarchical codebook with posterior matching (hiePM) \cite{Tara2019Active}:} 
This design also adopts the hierarchical codebook in \cite{alkhateeb2014channel}. However, unlike \cite{alkhateeb2014channel}, the proposed method in \cite{Tara2019Active} selects the beamforming vectors from the hierarchical beamforming codebook based on the AoA posterior distribution. It should be mentioned that this approach assumes that the fading coefficient value is known at the \changeWW{BS, so} that the AoA posterior distribution can be accurately computed. 

We first examine the performance of the proposed method and the above baselines when the AoA is uniformly drawn from a grid set of size $N=128$ with $\phi_\text{min}=-60^\circ$ and $\phi_\text{max}=60^\circ$. In this experiment, we consider that the BS is equipped with $M=64$ antennas. To fairly compare our proposed methods with other baselines, we set the number of uplink pilot transmission as $\tau=2\log_2(N)=14$. Further, we implement the proposed network on TensorFlow \cite{tensorflow2016} 
and Keras \cite{chollet2015} 
by employing Adam optimizer \cite{adam2014} with a learning rate progressively decreasing from $10^{-3}$ to $10^{-5}$. We consider $4$-layer neural networks with dense layers of widths $[1024,1024,1024,2M]$, respectively. For faster convergence, each dense layer is preceded by a batch normalization layer \cite{Ioffe2015}.
 We consider $10$ batches per epoch, each of which contains at most $2^{12}$ samples. To  investigate  the  ultimate performance of the proposed approach, we assume that we can generate as many data samples as needed for training the DNN.
We monitor the performance  of  the  DNN during  training  by  computing  the loss function for a validation data set of $10^5$ samples and  keep  the  model  parameters  that  have  achieved  the  best performance  so  far.  The  training  procedure is  terminated  when  the performance for the validation  set  has  not  improved  over $300$ epochs.

%%%%%%%% Fig. 7
 \begin{figure}[t]
 \centering
\includegraphics[width=0.49\textwidth]{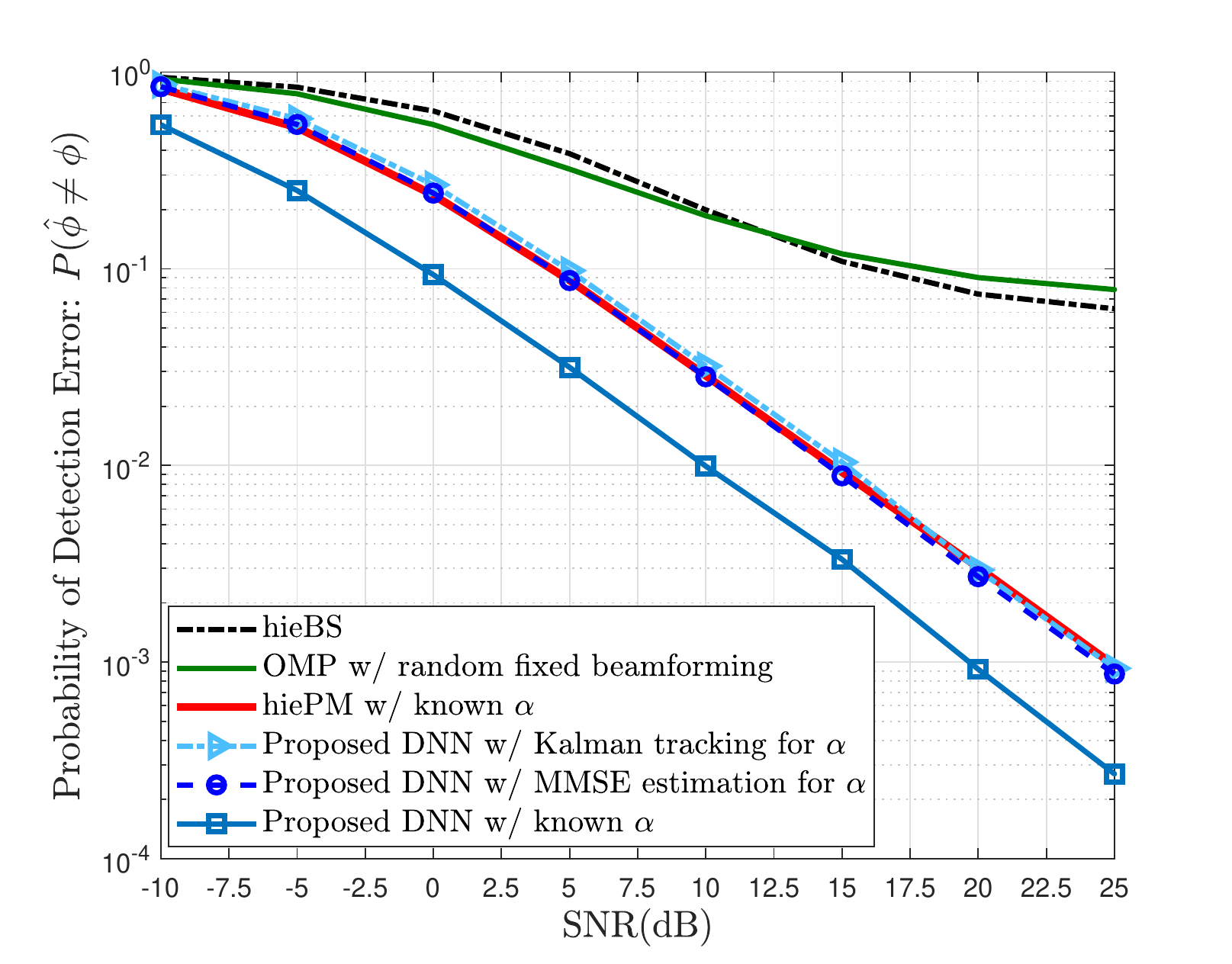}
\caption{Average detection error probability versus SNR for different methods in a system with $M=64$, $\tau=14$, and $\phi \in \{\phi_1,\phi_2,\ldots,\phi_N\}$ where $\phi_i = -60^\circ + \tfrac{i-1}{N-1} 120^\circ$. In this experiment, the sensing vectors of each method satisfy the \changer{$2$-norm constraint}.}
\label{fig:performance_detection}
\end{figure}

In Fig.~\ref{fig:performance_detection}, we plot the empirical error probability in AoA detection for different methods against \changeg{the} signal-to-noise-ratio, i.e., $\operatorname{SNR} \triangleq \log_{10}(P)$, over $10^5$ Monte Carlo trials. It can be seen that the OMP algorithm using random fixed beamforming and the hieBS algorithm both suffer from an error floor in the high SNR regime. However, the detection performance of  the proposed deep learning approach and the hiePM algorithm, both of which design the sensing vectors based on the AoA posterior distribution, continuously improves as the SNR increases. This observation \changeW{matches} the conclusion in \cite{Tara2019Active} that exploiting the measurement noise statistics via posterior matching can indeed improve the AoA detection performance. Furthermore, Fig.~\ref{fig:performance_detection} shows that the proposed approach for the known fading coefficient scenario can significantly outperform (by about $5$dB in this example) the hiePM algorithm in \cite{Tara2019Active}, which is developed based on the assumption that the fading component is given. This indicates that the proposed DNN can indeed design a better adaptive beamforming strategy as compared to the hiePM that employs the hierarchical codebook. \changeW{Further, we} can see that both estimation techniques proposed in Section \ref{sec:offgrid_unknown} to deal with the unknown fading coefficient scenario, i.e., MMSE estimation  and Kalman filtering, are indeed effective. In particular, the proposed DNN utilizing either of these two fading estimation techniques performs almost the same as hiePM, \changeW{which assumes the knowledge of}
the actual value of $\alpha$. \changeW{Note that since the Kalman filtering approach has lower storage complexity than the MMSE estimation approach, Kalman filtering would be preferable in practice.}

%%%%%%%% Fig. 8
 \begin{figure}[t]
 \centering
\includegraphics[width=0.49\textwidth]{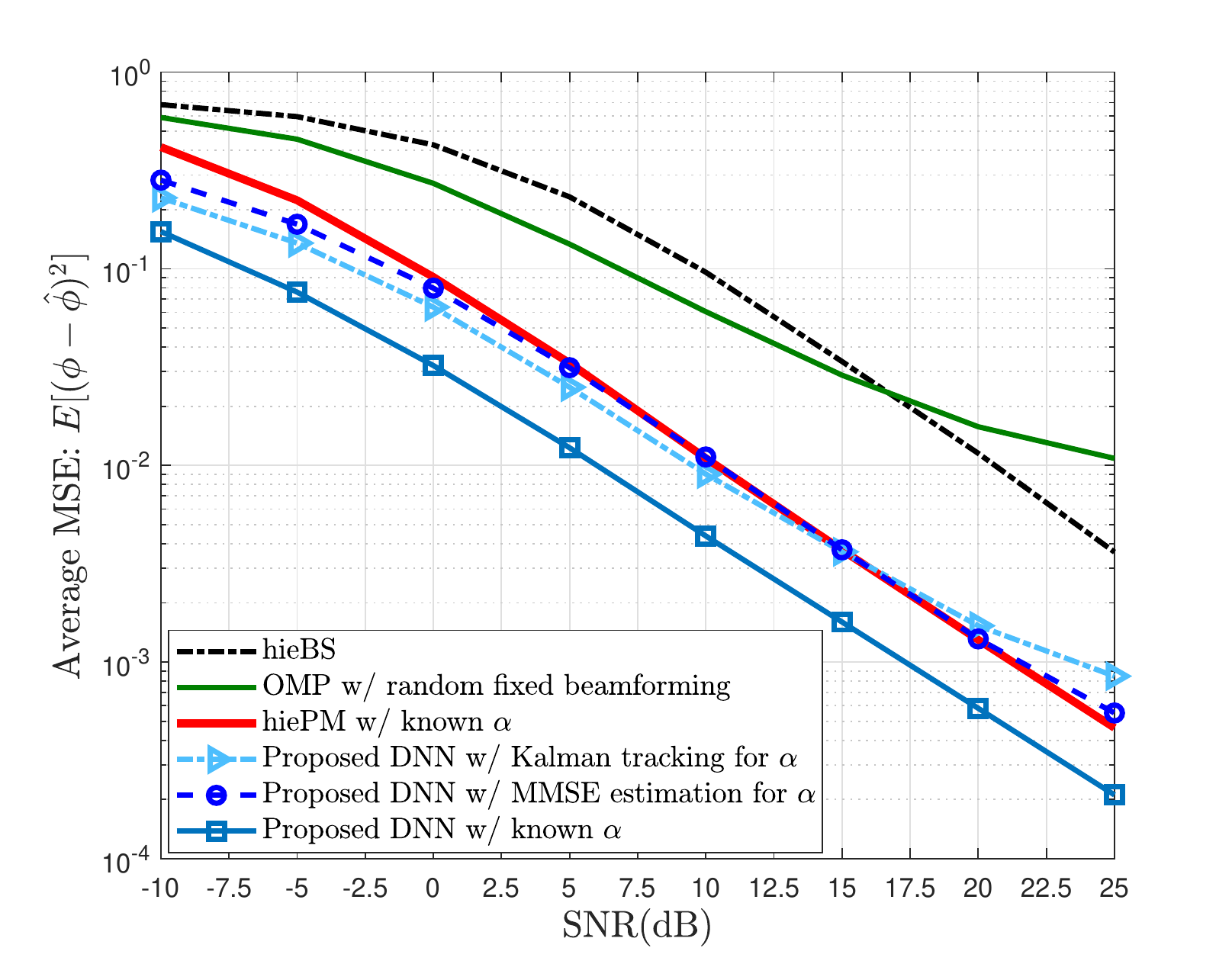}
\caption{Average MSE versus SNR for different methods in a system with $M=64$, $\tau=14$ and $\phi \in [-60^\circ,60^\circ]$. In this experiment, the sensing vectors of each method satisfy \changer{the $2$-norm constraint.}}
\label{fig:performance_estimation} 
\end{figure}

In the next experiment, we consider a more realistic setup where the AoA is a continuous variable, uniformly taken from the interval $[\phi_\text{min},\phi_\text{max}]=[-60^\circ,60^\circ]$. 
The other system parameters remain the same as in the previous experiment, i.e., $M=64$ and $\tau=14$. For the proposed method, we consider $N_c = 128$ posterior probability intervals and $N_c=20$ midpoint samples in each interval. For the baselines employing the hierarchical codebook, we use a hierarchical codebook with $\tfrac{\tau}{2} = 7$ stages consisting of $254$ sensing vectors. Further, for the OMP method and the hiePM algorithm, which have the on-grid AoA assumption, we set their grid size as $N=N_c\times N_s=2560$.
In Fig.~\ref{fig:performance_estimation}, we plot the empirical average MSE, computed over $10^5$ Monte Carlo runs, for different methods, \changeW{as functions of} SNR. Fig.~\ref{fig:performance_estimation} shows that, under the known fading coefficient assumption, there is a significant gap, i.e., about $5$dB, between the performance of the proposed DNN and that of the hiePM method. This indicates that the proposed strategy to deal with the AoA estimation problem for continuous-domain AoA is indeed effective. Further, we can see that the proposed DNN using either MMSE estimation or Kalman filter for estimating the fading coefficient achieves almost the same performance as the best baseline, which is the hiePM with the known fading assumption. This indicates that the proposed deep learning framework is superior to the state-of-the-art beamforming strategies for mmWave initial access. 
%\changeb{From Fig.~\ref{fig:performance_estimation}, we can also see that the Kalman tracking approach for $\alpha$ performs \changeg{slightly} better than the MMSE estimation. This phenomenon can probably be explained based on the nature of these two different approaches. The MMSE method considers a Gaussian prior distribution for the fading coefficient, i.e., $\alpha \sim \mathcal{CN}(0,1)$, and then seeks to find the estimate for $\alpha$ given $\theta$ based on the received signals \changeg{so far}. In contrast, the Kalman Filtering method is a two-level model \changeg{based} approach where the fading coefficient $\alpha$ is assumed to be Gaussian with an unknown mean and variance, then seeks to estimate the value of this mean and variance using the Kalman filter. We \changeg{believe} that in a low-SNR regime where the direct estimation of $\alpha$ may not be accurate, the two-level model \changeg{based} approach can be useful to capture statistics of the approximation error in the grid-less setup. In contrast, in a high SNR regime where a near-perfect estimation is possible via the MMSE estimator, modeling the fading coefficient with unknown mean and variance may lead to performance degradation.}

\begin{figure}[t]
 \centering
\includegraphics[width=0.37\textwidth]{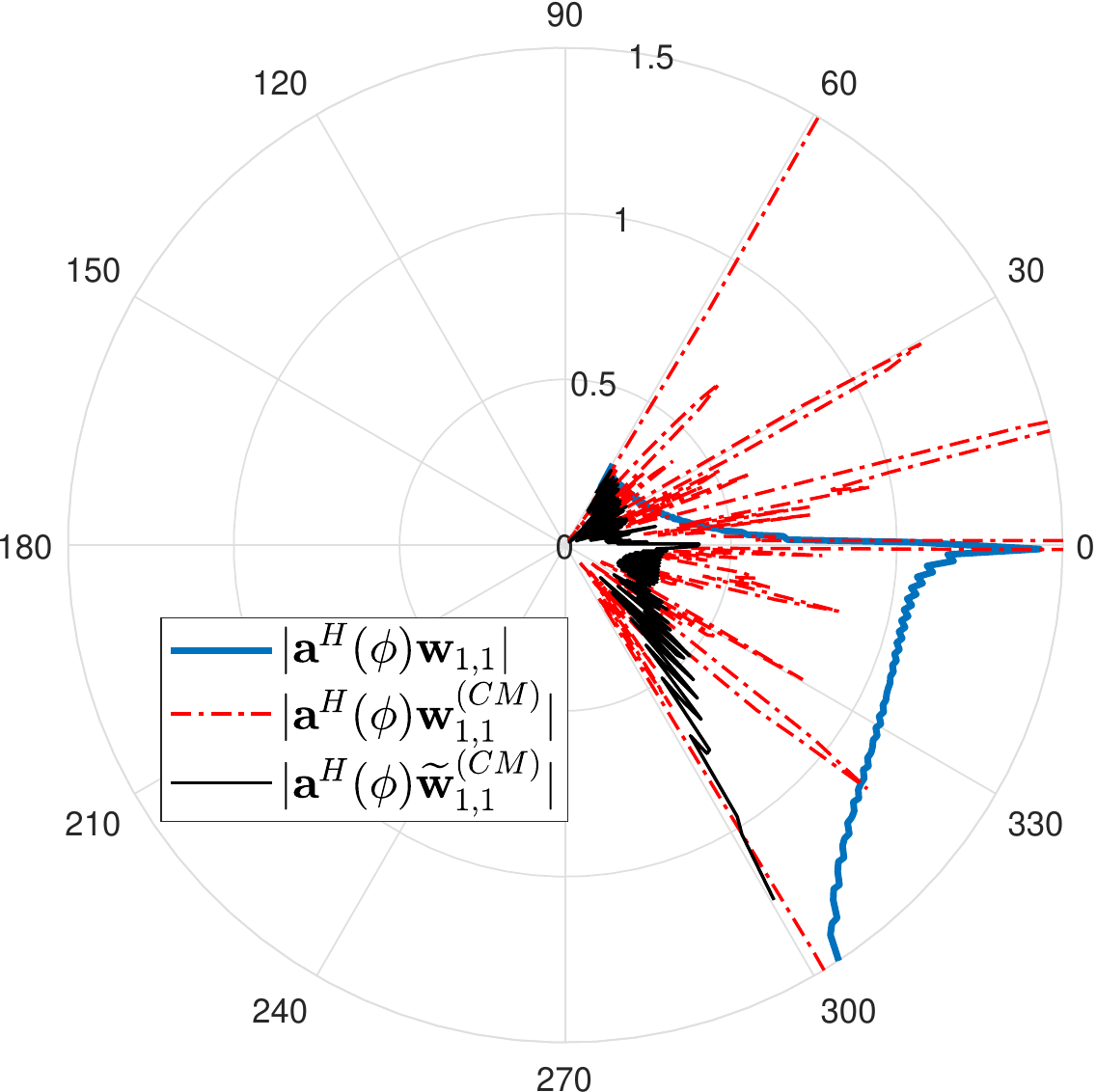}
\caption{The sensing filter of the first beamformer in stage one of the hierarchical codebook under the \changer{$2$-norm} constraint (i.e., the blue curve), and under the constant modulus norm constraint (i.e., the red curve based on the design in \cite{alkhateeb2014channel} and the black curve based on the algorithm in \cite{sohrabi2016hybrid}).}
\label{fig:hyb_filters} 
\end{figure}

As the final experiment, we consider a scenario where the analog sensing vectors are realized \changeg{using} a network of phase shifters, thus satisfying the constant modulus constraint. 
Accordingly, we need to enforce this constraint \changeg{for} the sensing vectors \changeg{in} all different methods. In the proposed DNN, we account for this constraint by employing a normalization layer as in \eqref{eq:activation_hyb}. For the OMP algorithm, we \changeW{can} simply generate a random matrix that satisfies such a constant modulus constraint. Accounting for the constant modulus constraint is, however, more challenging for methods using the hierarchical codebook. To deal with the constant modulus constraint in the hierarchical codebook, \cite{alkhateeb2014channel} suggests first \changeg{designing} relaxed version of each sensing vector that only satisfies \changeg{a $2$-norm} constraint, then approximate that relaxed design with the constant-modulus structure such that the Frobenius norm of the difference between the relaxed beamformer and the actual beamformer is minimized. For our problem of interest where the BS has only a single RF chain, by following the strategy of \cite{alkhateeb2014channel}, it can be shown that the $k$-th sensing vector in stage $s$ of the hierarchical codebook under constant modulus constraint can be written as: 
%37
\begin{equation}
\bw_{s,k}^{(\text{CM})} = \tfrac{1}{\sqrt{M}}e^{\imath \angle \bw_{s,k}},
\label{eq:hie_CM}
\end{equation}
where $\bw_{s,k}$ is the corresponding beamformer under the \changer{$2$-norm} constraint given by \eqref{eq:hie_total_power}. However, our numerical results show that the design in \eqref{eq:hie_CM} cannot preserve the hierarchical codebook's fundamental property, i.e., providing a constant beamforming gain for AoAs within the probing sector and nearly zero elsewhere. For example, Fig.~\ref{fig:hyb_filters}, \changelast{which} illustrates the shape of the sensing filter for $\bw_{s,k}$ and $\bw_{s,k}^{(\text{CM})}$, shows that while $\bw_{s,k}$ provides a nearly constant beamforming gain in the probing range and zero outside that region, $\bw_{s,k}^{(\text{CM})}$ fails to do so. 

%%%%%%%% Fig. 9

\changeW{To} design a better hierarchical codebook under the constant modulus constraint. \changeW{We} recognize that the solution for a hierarchical codeword in \eqref{eq:hie_total_power} is the least squares solution of the set of linear equations $\bA_\text{BS}^H \bw = \bg_{s,k}$. \changeg{Thus}, in order to obtain the corresponding sensing vector under the constant modulus constraint, we need to solve the following optimization problem:
%38
\begin{subequations}
\begin{align}
\label{eq:obj_hyb_filter}
\widetilde{\bw}_{s,k}^{(\text{CM})} = \argmin_{\bw} &~~~\| \bA_\text{BS}^H \bw - \bg_{s,k} \|^2_2\\
\text{s.t.}\hspace{9pt} &~~~|w_i| =\tfrac{1}{\sqrt{M}}, \forall{i}.
\end{align}
\end{subequations}
This is a challenging non-convex problem due to the constant modulus constraints. However, the decoupled nature of the constraints enables us to use the iterative coordinate descent algorithm in \cite{sohrabi2016hybrid}. In particular, it can be shown that the closed-form optimal solution for $w_i$ can be obtained when the other elements of the sensing vector are fixed.
%over the sensing vector elements (similar the idea in \cite{sohrabi2016hybrid}).
%By extracting the contribution of the $i$-th element of the sensing vector, the objective in \eqref{eq:obj_hyb_filter} can be written as the following form:
%\begin{equation}
%c_i + 2\Re\left({w_i^* \eta_i}\right),
%\end{equation}
%where $c_i$ and $\eta_i$ are both independent of $w_i$, e.g., $\eta_i = \mathbf{a}_i^H \bar{\bw}_i - f_i$ where $\mathbf{a}_i$ is the $i$-th column of $\bA_\text{BS}\bg_{s,k}$.
% if we assume that all elements of the sensing vector are fixed except $w_i$, the optimal value for $w_i$ is given by:
%\begin{equation}
%sa
%\end{equation}
This observation enables us to employ an iterative algorithm that starts with an initial feasible sensing vector, i.e., $\bw_{s,k}^{(\text{CM})} $, then sequentially updates each element of 
the sensing vector until the algorithm converges. Note that since in each \changeg{update step} of this algorithm, the objective function of \eqref{eq:obj_hyb_filter} decreases,  the convergence of the algorithm is guaranteed. It can be seen from Fig.~\ref{fig:hyb_filters}, that the above procedure can indeed improve the filter shape of the hierarchical codebook. However, the beamforming gain in the probing region under constant modulus constraint is still much smaller than \changeg{the} beamforming gain \changeg{without the constant modulus constraint}. This suggests that the performance of the hierarchical codebook-based methods may be significantly degraded under the constant modulus constraint.

%%%%%%%% Fig. 10
 \begin{figure}[t]
 \centering
\includegraphics[width=0.49\textwidth]{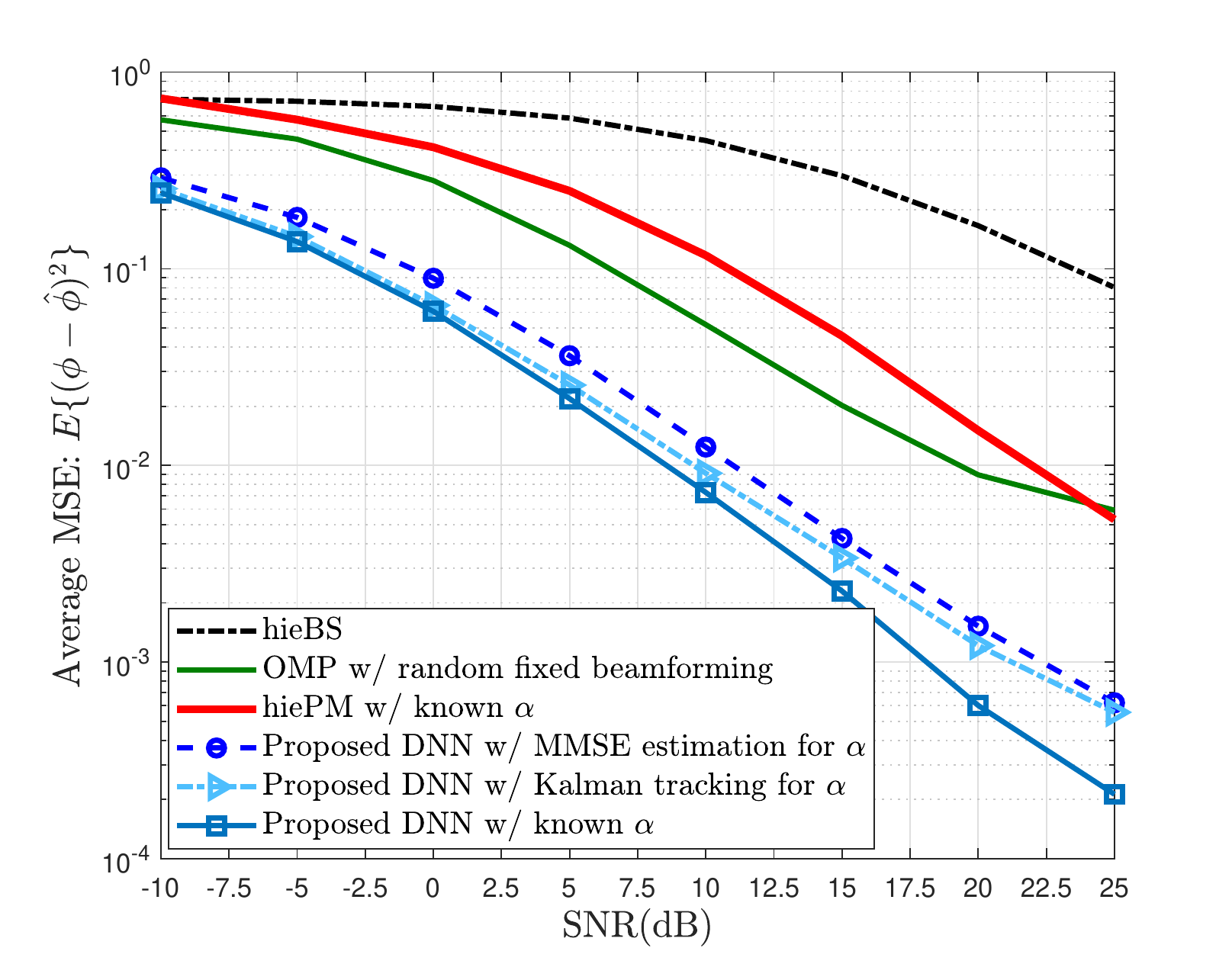}
\caption{Average MSE versus SNR for different methods in a system with $M=64$, $\tau=14$ and $\phi \in [-60^\circ,60^\circ]$. In this experiment, the entries of the sensing vectors of each method satisfy the constant modulus constraint.}
\label{fig:performance_estimation_hyb}
\end{figure}

We now evaluate the average MSE performance of different methods under the constant modulus constraint. Fig.~\ref{fig:performance_estimation_hyb} shows that even the OMP algorithm with fixed beamforming can outperform the hiePM and hieBS algorithms, both of which utilize the hierarchical codebook. This is aligned with our previous observation that an analog sensing vector implemented via simple analog phase shifters fails to accurately implement the idea of the hierarchical codebook. In contrast, we can see that the proposed DNN with or without the fading \changeg{coefficient} achieves excellent performance, indicating that the proposed deep learning framework can efficiently deal with practical constraints such as the constant modulus constraint.

\changeb{Finally, we provide a discussion about the computational complexity of the proposed deep learning framework as compared to the hiePM algorithm. We note that the computation comparison between the traditional optimization based methods and the proposed deep learning based approach is not straightforward since it is not clear how to account for the computations performed in the DNN training phase. Therefore, we only compare the computational complexity of the proposed deep learning approach in the operational phase with that of the hiePM algorithm. In this discussion, we consider a practical setup in which the fading coefficient $\alpha$ is not known a priori, hence both methods need to obtain the estimate for $\alpha$ using the Kalman filtering or MMSE estimation. Since both methods exploit the posterior distribution, their computational complexities for updating the posterior distribution are the same. As discussed in Section~\ref{sec:offgrid_unknown}, the computational cost of the posterior update in time frame $t$  is in the order of $O(t M N)$ if we use MMSE estimation or $O(M N)$ if we use the Kalman filter. For the adaptive beamforming design, the computational cost of the proposed DNN depends on the dimension and the number of layers adopted in the DNN architecture. Let $d_\ell$ denote the width of the $\ell$-th layer in an $L$-layer fully connected DNN. Then, the computational complexity of the forward propagation is in the order of $O\left(\sum_{\ell=0}^{L}  d_{\ell} d_{\ell+1}\right)$, but the practical run time of the DNN can be much less if it is implemented on a graphics processing unit (GPU). \changeF{Further, we remark that in the numerical experiments in this paper, the hidden layers' widths are set to be $d_\ell = 1024, ~ \forall \ell \in \{1,2,3\}$. However, we numerically observe that with a much narrower DNN with only $128$ neurons per hidden layer, almost the same performance can be achieved.}
To compare with hiePM, we expect 
%. On the other hand, the complexity of selecting the sensing vector based on the posterior distribution in the hiePM method depends on its hierarchical codebook size. An upper-bound of such computation is given by $O(2^S)$, where $S$ is the number of beam pattern levels in the hierarchical codebook. Based on this analysis, for typical scenarios, the computational complexity of designing the sensing vector based on the posterior distribution is higher in the proposed codebook-free approach as compared to the hiePM algorithm, i.e., $O\left(\sum_{\ell=0}^{L}  d_{\ell} d_{\ell+1}\right) \geq O(2^S)$. However, we expect
that the overall run time of the proposed DNN and that of hiePM to be in the same order since; i) the computational complexity of estimating the fading coefficient, which is common for both methods, is a non-negligible portion of the overall computational cost, ii) the computations needed for running the DNN in the operational phase can be highly parallelized which is not the case for the tree-search algorithm used in hiePM.}

%%%%%%%%%%%%%%%%%%%%%%%%%
%  		VII) Conclusion
%%%%%%%%%%%%%%%%%%%%%%%%%
\section{Conclusion}
\label{sec:conclusion}
This paper develops a deep learning framework for active learning of the angle-of-arrival of the channel's dominant path by adaptively designing the sequence of sensing vectors in the initial access phase of a mmWave communication. In particular, this paper proposes a DNN that directly maps the current AoA posterior distribution to the sensing vector of the next measurement. Further, to alleviate the computational burden of computing the AoA posterior distribution, this paper \changeFF{uses} an estimate of the fading coefficient to obtain an approximation of the AoA \changeFF{posterior distribution}.  This paper investigates two estimation strategies for the fading coefficient estimation, namely MMSE estimation and Kalman filter. This paper also proposes a modified DNN architecture to deal with the practical beamforming constant modulus constraint.  Numerical results indicate the proposed codebook-free deep \changeg{learning based} approach significantly outperforms the existing adaptive beam alignment techniques that utilize predesigned codebooks.

%%%%%%%%%%%%%%%%%%%%%%%%%%%%%%%%%%%%%
%%% References
%%%%%%%%%%%%%%%%%%%%%%%%%%%%%%%%%%%%%
\bibliographystyle{IEEEtran}
\bibliography{IEEEabrv,referenceF}

%%%%%%%%%%%%%%%%%%%%%%%%%%%%%%%%%%%%

%% Biography

%%%%%%%%%%%%%%%%%%%%%%%%%%%%%%%%%%%%

%%%%%%%%%%%%%%%%%% Foad Sohrabi
\begin{IEEEbiography}[{\includegraphics[width=1in,height=1.25in,clip,keepaspectratio]{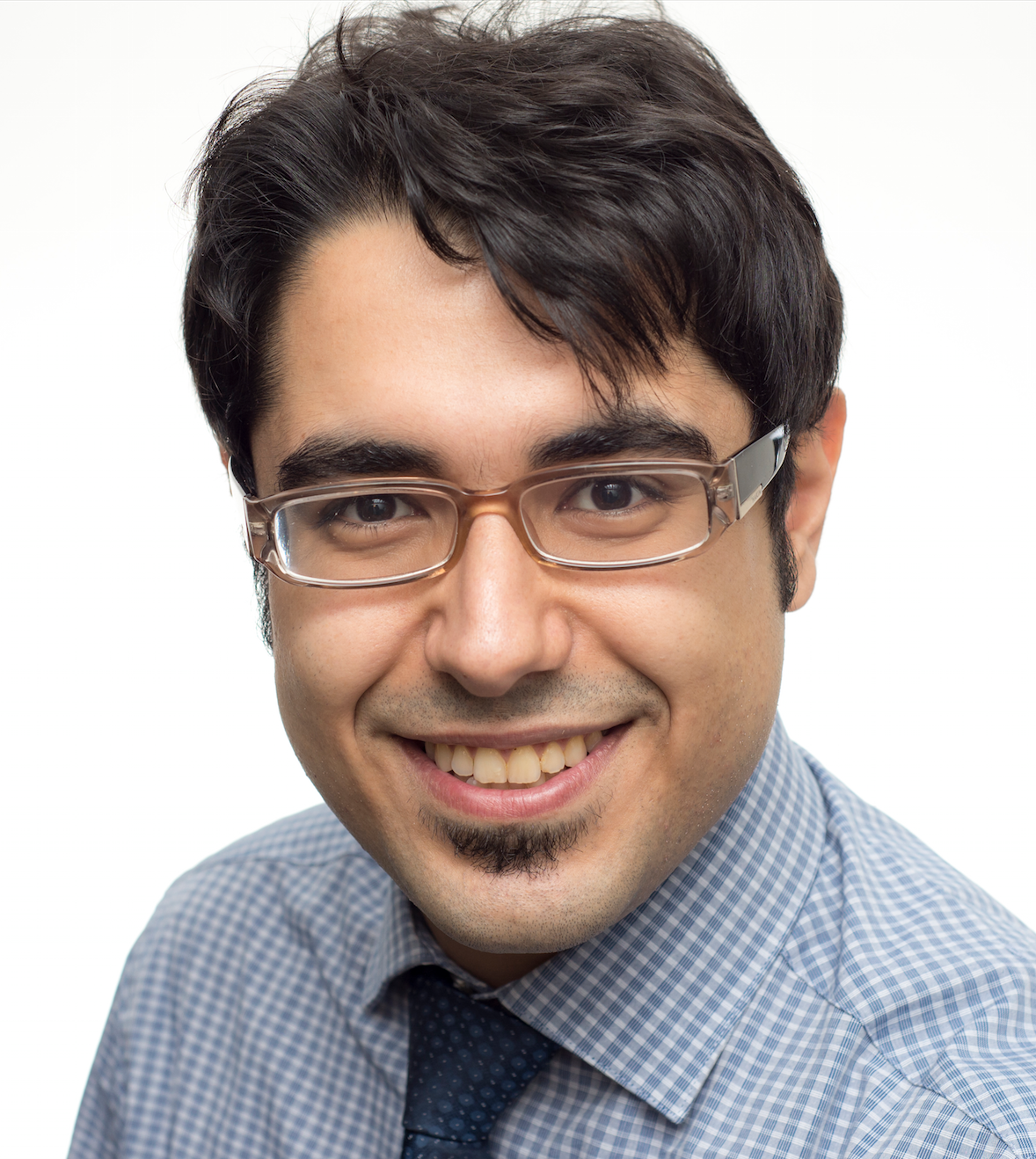}}]{Foad Sohrabi}
(S'13) received the B.A.Sc. degree from the University of Tehran, Tehran, Iran, in 2011, the M.A.Sc. degree from McMaster University, Hamilton, ON, Canada, in 2013, and the Ph.D. degree from the University of Toronto, Toronto, ON, Canada, in 2018, all in electrical and computer engineering. Since 2018, he has been a Post-Doctoral Fellow with the University of Toronto. In 2015, he was a Research Intern with Bell Labs, Alcatel-Lucent, Stuttgart, Germany. His research interests include MIMO communications, optimization theory, wireless communications, signal processing, and machine learning. He was a recipient of the IEEE Signal Processing Society Best Paper Award in 2017.
\end{IEEEbiography}

%%%%%%%%%%%%%%%%%% Zhilin Chen
\begin{IEEEbiography}[{\includegraphics[width=1in,height=1.25in,clip,keepaspectratio]{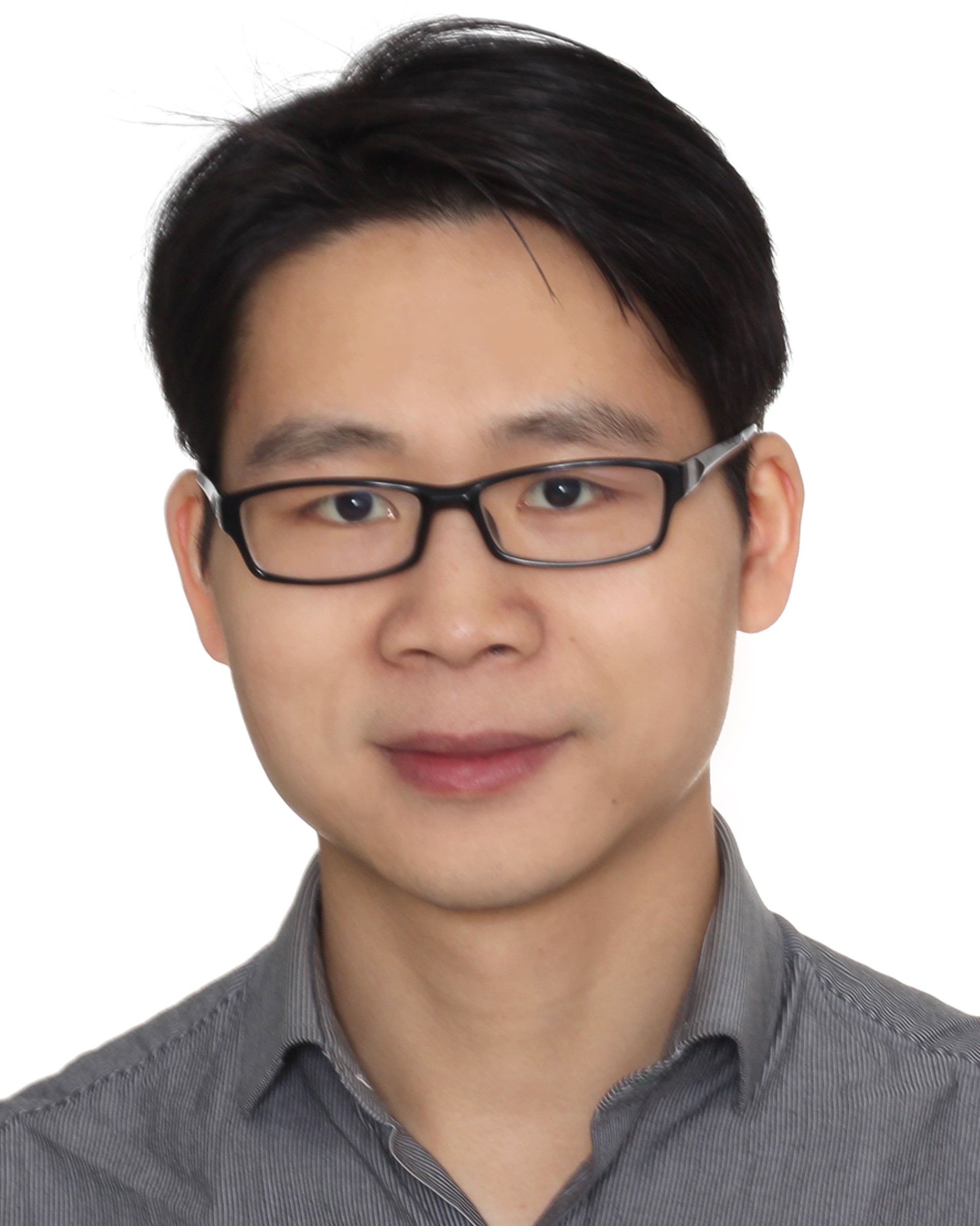}}]
{Zhilin Chen}
(S'14) received the B.E. degree in electrical and information engineering and the M.E. degree in signal and information processing from Beihang University (BUAA), Beijing, China, in 2012 and 2015, respectively, and the Ph.D. degree in electrical and computer engineering from the University of Toronto, Toronto, ON, Canada, in 2020. He is now with Huawei Technologies, Shenzhen, China. His main research interests include wireless communication, signal processing, and machine learning.
\end{IEEEbiography}

%%%%%%%%%%%%%%%%%%% We YU
\begin{IEEEbiography}[{\includegraphics[width=1in,height=1.25in,clip,keepaspectratio]{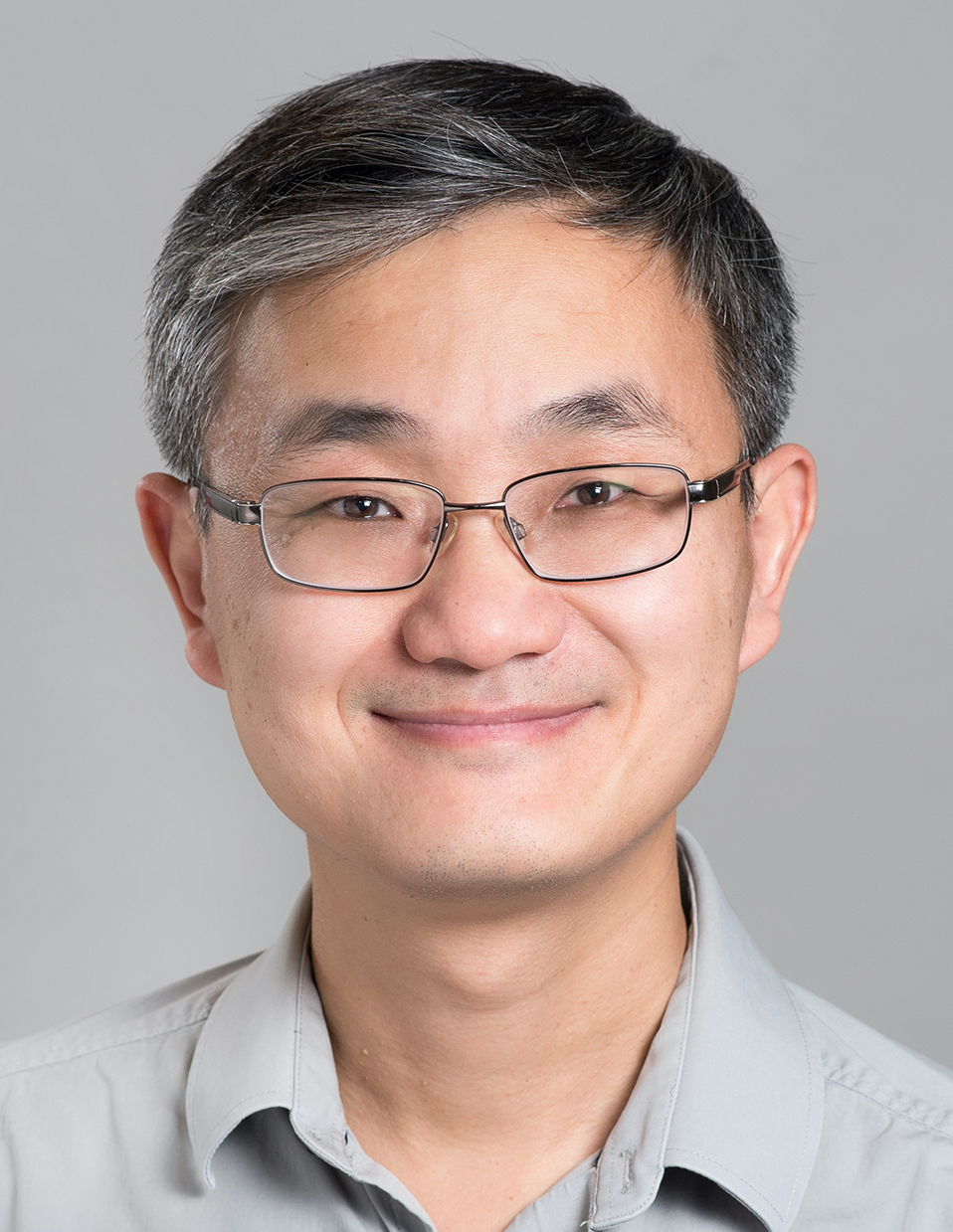}}]
{Wei Yu} (Fellow, IEEE) received the B.A.Sc. degree in computer engineering and mathematics from the University of Waterloo, Waterloo, ON, Canada, in 1997, and the M.S. and Ph.D. degrees in electrical engineering from Stanford University, Stanford, CA, USA, in 1998 and 2002, respectively. Since 2002, he has been with the Electrical and Computer Engineering Department, University of Toronto, Toronto, ON, Canada, where he is currently a Professor and holds the Canada Research Chair (Tier 1) in Information Theory and Wireless Communications. Prof. Wei Yu is the President of the IEEE Information Theory Society in 2021, and has served on its Board of Governors since 2015. He is a Fellow of the Canadian Academy of Engineering and a member of the College of New Scholars, Artists, and Scientists of the Royal Society of Canada. He received the Steacie Memorial Fellowship in 2015, the IEEE Marconi Prize Paper Award in Wireless Communications in 2019, the IEEE Communications Society Award for Advances in Communication in 2019, the IEEE Signal Processing Society Best Paper Award in 2017 and 2008, the Journal of Communications and Networks Best Paper Award in 2017, and the IEEE Communications Society Best Tutorial Paper Award in 2015. He served as the Chair of the Signal Processing for Communications and Networking Technical Committee of the IEEE Signal Processing Society from 2017 to 2018. He was an IEEE Communications Society Distinguished Lecturer from 2015 to 2016. He is currently an Area Editor of the IEEE Transactions on Wireless Communications, and has in the past served as an Associate Editor for IEEE Transactions on Information Theory, IEEE Transactions on Communications, and IEEE Transactions on Wireless Communications. 
\end{IEEEbiography}

\end{document}